\documentclass[aps,pre,reprint,showpacs,amsmath,amssymb,final,letterpaper]{revtex4-1}

\usepackage{txfonts}
\usepackage{graphicx}
\usepackage{subfigure}
\usepackage{wasysym}
\usepackage[pdftex]{hyperref}
\usepackage[usenames,dvipsnames]{color}
\usepackage{array}

\begin{document}


\title{Modeling the dynamical interaction between epidemics on overlay networks}

\author{Vincent Marceau}
\author{Pierre-Andr\'e No\"el}
\author{Laurent H\'ebert-Dufresne}
\author{Antoine Allard}
\author{Louis J. Dub\'e}
\homepage{http://dynamica.phy.ulaval.ca}
\affiliation{D\'epartement de Physique, de G\'enie Physique, et d'Optique, Universit\'e Laval, Qu\'ebec, Qu\'ebec, Canada G1V 0A6}

\date{\today}

\begin{abstract}
Epidemics seldom occur as isolated phenomena. Typically, two or more viral agents spread within the same host population and may interact dynamically with each other. We present a general model where two viral agents interact via an immunity mechanism as they propagate simultaneously on two networks connecting the same set of nodes. Exploiting a correspondence between the propagation dynamics and a dynamical process performing progressive network generation, we develop an analytic approach that accurately captures the dynamical interaction between epidemics on overlay networks. The formalism allows for overlay networks with arbitrary joint degree distribution and overlap. To illustrate the versatility of our approach, we consider a hypothetical delayed intervention scenario in which an immunizing agent is disseminated in a host population to hinder the propagation of an undesirable agent (e.g. the spread of preventive information in the context of an emerging infectious disease). 
\end{abstract}


\maketitle


\section{Introduction \label{sec:intro}}

Epidemic dynamics has been largely studied with the help of mathematical models in which a single viral agent propagates in a given host population. Although the paradigm of isolated epidemics may be well suited in some cases, there are numerous other situations in which more than one process occurs and interacts in the same population. Different biological pathogens may interact through ecological~\cite{rohani98_prslb,vasco07_jtb} and immunological~\cite{abu-raddad06_science,vasco07_jtb} mechanisms, or multiple strains of the same disease may compete for hosts according to some cross-immunity profile~\cite{andreasen97_jmb,kamo02_physica,abu-raddad08_prl}. The spread of fear or awareness in the context of an emerging disease~\cite{bagnoli07_pre,epstein08_plos,funk09_pnas,funk10_jtb} can also be considered as a case of interacting viral agents, i.e., information and disease. In computer networks, the dissemination of countermeasures using a contagious vaccination scheme has been suggested to counter harmful computer viruses~\cite{chen04_ieee,goldenberg05_nature,stauffer06_pre}. 

When propagating in some host population, two viral agents may follow different -- or share similar -- routes of transmission. Taking into account how individuals are in contact with each other then becomes of great importance when modeling their interaction. By explicitly considering those heterogeneous contact patterns between individuals, \emph{network-based} models are an ideal framework for the study of interacting epidemics in structured populations~\cite{keeling05_jrsi,meyers07_bams}. 

The interaction between two viral agents has been studied from the perspective of complex networks in a limited number of contributions~\cite{chen04_ieee,goldenberg05_nature,newman05_prl,ahn06_pre,stauffer06_pre,bansal09_arxiv,funk09_pnas,bansal10_plos,funk10_jtb,funk10_pre}.  An important step towards a general theory of interacting processes on complex networks was recently accomplished by Funk and Jansen~\cite{funk10_pre}. Generalizing the previous work of Newman~\cite{newman05_prl}, they analyzed the interaction between two viral agents propagating \emph{successively} on overlay networks, i.e., two networks connecting the same set of nodes. Albeit very elegant, their analytical approach, based on bond percolation, is static and does not apply to the case of \emph{dynamically interacting} viral agents. 

The purpose of this contribution is to develop an analytical approach able to capture the dynamical interaction between viral agents spreading \emph{simultaneously} on overlay networks. To this end, we make use of a correspondence between propagation on networks and a dynamical process performing progressive network generation~\cite{noel11_arxiv} (see also Appendix C of~\cite{noel09_pre}). The formalism obtained is quite general, and allows for overlay networks with arbitrary joint degree distribution and overlap. The language of epidemiology is used throughout this work for its clarity, yet our approach may be applied to processes of other nature that spread on networks.

This paper is organized as follows. In Sec.~\ref{sec:model}, we introduce a model in which two viral agents propagate and interact dynamically on overlay networks. The analytical approach is then developed in two steps in Sec.~\ref{sec:formalism}. Some properties of the model are investigated in Sec.~\ref{sec:results}, where we also validate the accuracy of the analytical predictions by comparison with Monte Carlo simulations of the dynamics. Our conclusions are summarized in Sec.~\ref{sec:conclusion}. An online supplementary document~\footnote{See EPAPS Document No. [number will be inserted by publisher] for the full set of equations of our analytical approach.} containing the full set of equations of our analytical approach completes this contribution.

\section{Interacting epidemics on overlay networks \label{sec:model}}

In this section, we introduce a general epidemic model where two dynamically interacting viral agents spread on overlay networks.

\subsection{Overlay networks \label{sec:overlay}}

\begin{figure}[!t]
\subfigure[]{\includegraphics[width = .45\columnwidth]{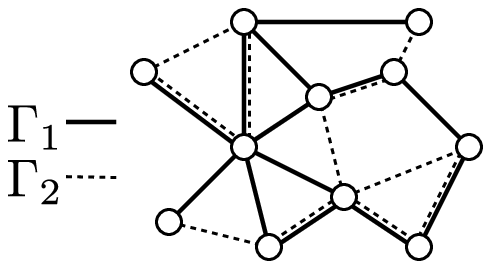} \label{fig:networks2a}} \qquad
\subfigure[]{\includegraphics[width = .45\columnwidth]{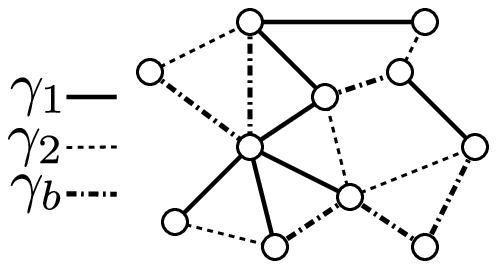} \label{fig:networks2b}}
\caption{ (a) Schematical illustration of a system of two overlay networks, $\Gamma_1=(V, E_1)$ and $\Gamma_2=(V, E_2)$. (b) This system can be decomposed into three non-overlapping networks: $\gamma_1=(V, E_1 \setminus E_2)$, $\gamma_2=(V, E_2 \setminus E_1)$, and $\gamma_b=(V, E_1 \cap E_2)$. \label{fig:networks} }
\end{figure}

A system of two \emph{overlay networks} is defined by two networks, $\Gamma_1=(V,E_1)$ and $\Gamma_2=(V,E_2)$, which connect the same set of \emph{nodes} $V$ through their own set of \emph{links}, $E_1$ and $E_2$ [see Fig.~\ref{fig:networks2a}]~\cite{funk10_pre}. Nodes represent individuals of a given \emph{host population}, while links correpond to potential transmission routes between pairs of individuals. As usually done in the network litterature, we denote the size of the host population by $N \equiv |V|$. Two nodes are said to be \emph{neighbors} on $\Gamma_g$, where $g\in \{1,2\}$, if they share a link on this network, and the number of neighbors of a node on $\Gamma_g$ define its \emph{degree} $k_g$ on this network. Neither can a node be linked to itself (no self-loops) nor share more than one link with another node on the same network (no repeated links). 

The system is characterized by its \emph{joint degree distribution} $P(k_1,k_2) \equiv P(\boldsymbol{k})$, which corresponds to the probability that a node selected at random in the host population has a degree $k_1$ on $\Gamma_1$ and $k_2$ on $\Gamma_2$. The \emph{marginal degree distribution} $P_g(k_g)$ of each network can be obtained by summing over $P(\boldsymbol{k})$, i.e., 
\begin{align}
P_1(k_1)=\sum_{k_2}P(k_1,k_2) \ \ , \ \ P_2(k_2)=\sum_{k_1}P(k_1,k_2) \ .
\end{align}

Even if $E_1 \neq E_2$ in the general case, a given number of links may be common to both networks. Because $P(\boldsymbol{k})$ contains no information about the potential \emph{overlap} in the system, we will resort to a useful decomposition into three non-overlapping networks~\cite{funk10_pre}. Let $\gamma_1=(V, E_1 \setminus E_2)$ be the network characterized by all the links unique to $\Gamma_1$, $\gamma_2=(V, E_2 \setminus E_1)$ be the network characterized by all the links unique to $\Gamma_2$, and $\gamma_b=(V, E_1 \cap E_2)$ be the network containing all links common to $\Gamma_1$ and $\Gamma_2$ [see Fig.~\ref{fig:networks2b}]. Because it contains information about the distribution of overlapping links, the joint degree distribution resulting from the three-networks decomposition, denoted by $\rho(\kappa_1,\kappa_2,\kappa_b) \equiv \rho(\boldsymbol{\kappa})$, offers a more accurate description of the system. The distribution $P(\boldsymbol{k})$ can be obtained from $\rho(\boldsymbol{\kappa})$ by summing over $\kappa_b$,
\begin{align}
P(\boldsymbol{k}) = \sum_{\kappa_b} \rho(k_1-\kappa_b,k_2-\kappa_b,\kappa_b) \ .
\end{align}

\subsection{Interacting epidemics \label{sec:SIR}}

In the single viral agent \emph{susceptible-infectious-recovered} (SIR) dynamics on a network, nodes are divided into three states: susceptible ($S$), infectious ($I$), or recovered ($R$). \emph{Infectious contacts} occur between infectious nodes and their neighbors at the rate $\beta$. If a susceptible node is involved in an infectious contact, \emph{transmission} ensues and it becomes infectious. Infectious nodes recover at the rate $\alpha$, and become immune to further infection.

Here, we extend the SIR dynamics to two interacting viral agents. We assume that ``agent 1'' propagates via the links of $\Gamma_1$, while ``agent 2'' spreads on $\Gamma_2$. If both viral agents are transmitted through the same type of interactions between individuals, then $\Gamma_1=\Gamma_2$; otherwise, $\Gamma_1\neq \Gamma_2$. 

At any time, the state of a node is given by the combination of its particular state regarding each viral agent. For agent $g$, its \emph{g-state} can be either \emph{g-susceptible} ($S_g$), \emph{g-infectious} ($I_g$), or \emph{g-recovered} ($R_g$). The rate of infectious contacts on $\Gamma_g$ is $\beta_g$, while $g$-infectious nodes recover at the rate $\alpha_g$ and become immune to further infection by agent $g$. In order to study the dynamical interaction between epidemics, we consider a case of \emph{leaky partial immunity}~\cite{bansal09_arxiv}. When an infectious contact occurs between a $g$-infectious node and a $g$-susceptible node whose state regarding the other agent $\hat{g}$ is $Y_{\hat{g}} \in \{ S_{\hat{g}}, I_{\hat{g}}, R_{\hat{g}} \}$, transmission of agent $g$ successfully follows with probability $\sigma_g^Y$ and the $g$-susceptible node becomes $g$-infectious. Otherwise, it remains $g$-susceptible with complementary probability $\overline{\sigma}_g^Y \equiv 1-\sigma_g^Y$. We assume that an infectious contact on $\Gamma_g$ may occur only once between two given nodes. 

The motivation for this interaction rule is that, while remaining general, it renders the model analytically tractable using a reasonable level of complexity. Other interaction mechanisms could have been considered, such as perfect partial immunity~\cite{bansal09_arxiv} or leaky partial immunity allowing for more than one infectious contact between two nodes. More on this topic is covered in Sec.~\ref{sec:other}. 

\subsection{Monte Carlo simulations \label{sec:MC}}

Monte Carlo simulations of epidemic propagation on overlay networks are performed in two steps, network generation and viral agent propagation. Our overlay networks are generated using two different algorithms, both adapted from the well-known \emph{configuration model}~\cite{molloy95_rsa,newman02_pre}.

The first algorithm, based on the joint degree distribution $P(\boldsymbol{k})$, generates two overlay networks with \emph{random overlap}. (i) A random degree sequence $\{ \boldsymbol{k}_i \}$ of length $N$ subjected to $P(\boldsymbol{k})$ is generated. Since a link consists of two \emph{stubs}, we ensure that $\sum_i k_{i,g}$ is even for all $g\in \{1,2\}$, otherwise an element of the degree sequence is selected at random and generated again; (ii) For each $\boldsymbol{k}_i$, a node with $k_{i,1}$ stubs on $\Gamma_1$ and $k_{i,2}$ stubs on $\Gamma_2$ is created; (iii) Independently for each network, pairs of unconnected stubs are randomly chosen and connected together until all unconnected stubs are exhausted; (iv) The presence of self-loops and repeated links is tested on each network. All faulty links on a network are removed by randomly choosing a pair of connected stubs on the same network and rewiring them to the former stubs. 

To generate networks with \emph{arbitrary overlap}, we use a second algorithm based on the joint degree distribution $\rho(\boldsymbol{\kappa})$. The procedure is essentially the same as described above, except that three networks ($\gamma_1$, $\gamma_2$, and $\gamma_b$) are generated from $\rho(\boldsymbol{\kappa})$ with the additional constraint that \emph{one link cannot exist on more than one network}. The networks $\Gamma_1$ and $\Gamma_2$ are then constructed from $\gamma_1$, $\gamma_2$, and $\gamma_b$ as illustrated in Fig.~\ref{fig:networks}.

The viral agent propagation phase is carried out using discrete time steps of length $\Delta t$. At each time step and for viral agent $g \in \{1,2\}$, every link on $\Gamma_g$ between a $g$-infectious and a $g$-susceptible node is tested for infectious contacts with probability $\beta_g \Delta t$, on the condition that it did not happen previously. If the test returns positive, the $g$-susceptible node becomes $g$-infectious with transmission probability $\sigma_g^Y$, where $Y_{\hat{g}}$ corresponds to its state regarding the other agent $\hat{g}$. Recovery events are tested with probability $\alpha_g \Delta t$. All simulations are initialized by infecting at random a fraction $\epsilon_1$ of the nodes in the system with agent 1 at time $t=0$. To allow for a delay between epidemics, the dynamics of agent 2 are initialized at time $t=\tau\geq0$ with the random infection of a fraction $\epsilon_2$ of the nodes in the system. 

The simulations presented in this paper are carried out on networks of size $N=25\, 000$ (unless explicitly noted) with $\Delta t=0.001$. Both recovery rates are set to unity, $\alpha_1=\alpha_2=1$. Moreover, we will use $\epsilon_1=\epsilon_2=0.001$ as initial conditions. All Monte Carlo results shown in the figures are computed over a total of $1000$ simulations unless explicitly noted.

\section{Mean-field approach \label{sec:formalism}}

We now develop a network-based compartmental formalism that describes the dynamics of the model introduced in the previous section. Our approach is based on the concept of \emph{on the fly network generation} recently introduced in~\cite{noel11_arxiv}. The case of interacting epidemics on overlay networks with random overlap is treated in the first place, and the approach is later generalized to allow for arbitrary overlap. Finally, we provide some insights on how to handle similar types of interaction dynamics. 

\subsection{Networks with random overlap \label{sec:random}}

In order to develop the analytical approach that follows, some preliminary considerations are needed. In Sec.~\ref{sec:MC}, we explained the two-step procedure used in Monte Carlo simulations of epidemics on configuration model networks. However, as we recently pointed out in a recent contribution~\cite{noel11_arxiv}, an alternative procedure, consisting of one single step where the networks are generated on the fly (i.e., during propagation, when required), can be considered.

\begin{figure}[t]
\mbox{
\subfigure[\ $t=t_0$ (initial state)]{\includegraphics[width = .35\columnwidth]{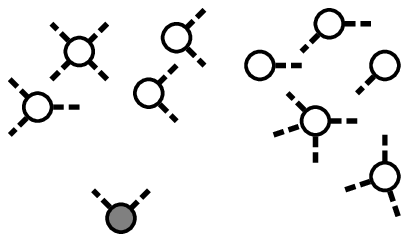} \label{fig:otf1}} \qquad
\subfigure[\ $t=t_1$]{\includegraphics[width = .35\columnwidth]{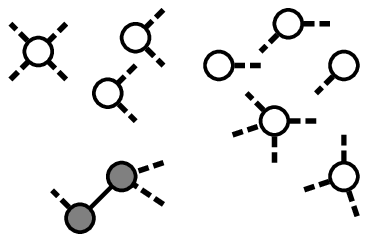} \label{fig:otf2}}} \\
\mbox{
\subfigure[\ $t=t_2$]{\includegraphics[width = .35\columnwidth]{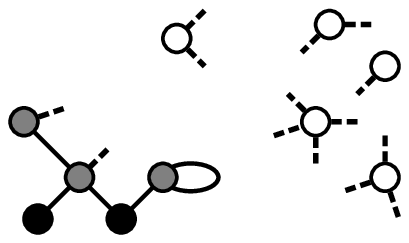} \label{fig:otf3}} \qquad
\subfigure[\ $t=t_3$ (final state)]{\includegraphics[width = .35\columnwidth]{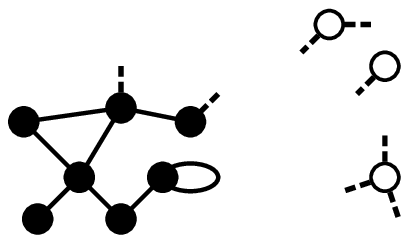} \label{fig:otf4}}} 
\caption{Schematical illustration of on the fly network generation for a single viral agent susceptible-infectious-removed epidemic (\Circle : susceptible nodes, {\color{Gray}\CIRCLE}: infectious nodes, \CIRCLE : recovered nodes).  (a) Initially, one node of degree 2 is infectious. (b) When an infectious contact occurs over one stub belonging to the infectious node, it is matched with another stub chosen at random between all the unmatched stubs (dashed lines), thus forming a link (solid lines). If previously susceptible, the assigned neighbor becomes infectious. (c) Stubs are progressively matched pairwise as infectious contacts occur in the population. In the matching process, self-loops and repeated links are allowed, but their probability decreases as $N^{-1}$. (d) The epidemic stops when there are no more infectious nodes. Parts of the network not reached by the viral agent are never built. \label{fig:onthefly}}
\end{figure}

Recall that in the procedure of Sec.~\ref{sec:MC}, we test for infectious contact \emph{every link} on $\Gamma_g$ between a $g$-susceptible and a $g$-infectious node, on the condition that an infectious contact never happened between them. Suppose instead we used to test for infectious contact \emph{every stub} on $\Gamma_g$ that emanates from a $g$-infectious node and that has never been the host of an infectious contact, without any further distinction. Then \emph{only} when this test returned positive would we have inquired about the state of the corresponding neighbor. If $g$-susceptible, then we would have tested for transmission. Unlike in Sec.~\ref{sec:MC}, this new procedure does not require any explicit prior knowledge of the structure of the networks in the system: neighbors can be assigned on the fly by matching stubs pairwise at the moment infectious contacts occur. A schematization of on the fly network generation for a single agent SIR epidemic is illustrated in Fig.~\ref{fig:onthefly}. 

In the case of overlay networks with random overlap, i.e., when stubs are matched independently at random on $\Gamma_1$ and $\Gamma_2$, an almost exact correspondence may be established between the algorithm exposed in Sec.~\ref{sec:MC} and an equivalent stochastic Markov process performing on the fly network generation~\cite{noel11_arxiv}. The only difference arises from self-loops and repeated links, which are allowed in the stochastic process. However, since their probability decreases as $N^{-1}$, we expect the results to agree in the limit of large networks.

Instead of tracking the full stochastic process, we rather focus on mean values in the asymptotic limit $N\to\infty$ and hence obtain a fully deterministic approach. Let $[X_1Y_2]_{ij}(t)$ be the mean fraction of nodes in the system that are of 1-state $X_1$, of 2-state $Y_2$, have $i$ unmatched stubs on $\Gamma_1$, and $j$ unmatched stubs on $\Gamma_2$ at time $t$~\footnote{For the sake of readability, we will drop from here on the explicit time dependence when obvious.}. The ordinary differential equation (ODE) governing the time evolution of the $[X_1Y_2]_{ij}$ compartment consists of two parts, accounting for the dynamics of each viral agent. As both parts are very similar for corresponding states (e.g., agent 1 dynamics for 1-susceptible nodes and agent 2 dynamics for 2-susceptible nodes), we exclusively focus on agent 1 dynamics. The same considerations apply to agent 2 dynamics as well. 

Let $\Theta_1$ be the probability that an unmatched stub on $\Gamma_1$ belongs to a 1-infectious node,
\begin{align}
\Theta_1 = \dfrac{\sum_Y \sum_{i,j} i[I_1Y_2]_{ij}}{\sum_{X'\!,Y'} \sum_{i'\!,j'} i'[X'_1Y'_2]_{i'j'}} \ .
\label{eq:Theta1}
\end{align}
Nodes in the $[S_1Y_2]_{ij}$ compartment will lose unmatched stubs on $\Gamma_1$ at the rate $\beta_1 \Theta_1 i$ as they are involved in infectious contacts with 1-infectious individuals. When this happens, they are either transferred to the $[I_1Y_2]_{(i-1)j}$ compartment with probability $\sigma_1^Y$ (successful transmission), or to the $[S_1Y_2]_{(i-1)j}$ compartment with probability $\overline{\sigma}_1^Y$ (failed transmission). This yields the following contribution to the ODE governing $[S_1Y_2]_{ij}$:
\begin{align}
 \beta_1 \Theta_1 \! \left[ (i\! +\! 1)\overline{\sigma}_1^Y \! [S_1Y_2]_{(i+1)j} \! - i[S_1Y_2]_{ij} \right] \ .
\label{eq:dSYdtNO}
\end{align}

Nodes are transferred from $[I_1Y_2]_{ij}$ to $[R_1Y_2]_{ij}$ at the rate $\alpha_1$ due to recovery events. Moreover, a node from $[I_1Y_2]_{ij}$ can be transferred to $[I_1Y_2]_{(i-1)j}$ due to a loss of an unmatched stub, which may occur in two different ways: if it is the source of an infectious contact (at rate $\beta_1 i$), or if it is victim of an infectious contact originating from another 1-infectious node (at rate $\beta_1\Theta_1 i$). Adding the incoming flow of newly infectious 1-susceptible nodes, one obtains the following contribution of agent 1 dynamics to the ODE governing $[I_1 Y_2]_{ij}$:
\begin{equation}
\begin{gathered}
- \alpha_1 [I_1Y_2]_{ij} + \beta_1 \Theta_1 (i\! +\! 1)\sigma_1^Y [S_1Y_2]_{(i+1)j} \\
+ \beta_1(1\! + \! \Theta_1) \! \left[ (i\! +\! 1)[I_1Y_2]_{(i+1)j} \! - i[I_1Y_2]_{ij} \right] \ .
\label{eq:dIYdtNO}
\end{gathered}
\end{equation}

Nodes are removed from $[R_1Y_2]_{ij}$ and transferred to $[R_1Y_2]_{(i-1)j}$ at the rate $\beta_1 \Theta_1 i$ as they are involved in infectious contacts with 1-infectious individuals. Including the incoming flow of newly recovered 1-infectious nodes, the contribution of agent 1 dynamics to the ODE governing $[R_1 Y_2]_{ij}$ reads
\begin{align}
\alpha_1 [I_1Y_2]_{ij} + \beta_1 \Theta_1 \! \left[ (i\! +\! 1)[R_1Y_2]_{(i+1)j} \! - i[R_1Y_2]_{ij} \right] \ .
\label{eq:dRYdtNO}
\end{align}

Finally, in order for the dynamics to be completely specified, an initial condition is required for each compartment. As mentioned earlier, random infection of a fraction $\epsilon_1$ of the population with agent 1 occurs at $t=0$, which gives:
\begin{align}
[X_1Y_2]_{ij} (0) = 
\begin{cases} 
(1\! -\! \epsilon_1)P(i,j) & \textrm{if} \ X\! =\! S \ \textrm{and} \ Y\! =\! S \\
\epsilon_1P(i,j) & \textrm{if} \ X\! =\! I \ \textrm{and} \ Y\! =\! S \\
0 & \textrm{otherwise} \ .
\end{cases} 
\end{align}
Agent 2 is then introduced at random in the population at time $t=\tau$ and infects a fraction $\epsilon_2$ of the nodes. This is implemented by making the following substitution at $t=\tau$:
\begin{align}
[X_1Y_2]_{ij} (\tau) \to
\begin{cases} 
(1\! -\! \epsilon_2)[X_1 S_2]_{ij}(\tau) & \textrm{if} \ Y\! =\! S \\
\epsilon_2[X_1S_2]_{ij}(\tau) & \textrm{if} \ Y\! =\! I \\
0 & \textrm{otherwise} \ .
\end{cases} 
\end{align}

The full system of ODEs describing the dynamics of the model on overlay networks with random overlap can be found in the supplementary document of this paper~\footnotemark[1]. 

\subsection{Networks with arbitrary overlap \label{sec:arbitrary}}

Generalization to the case of arbitrary overlap requires additional considerations. They originate from the fact that the two networks are not built independently anymore by matching stubs at random on each network. Since the structure of one network is now influenced by the structure of the other, an exact correspondence cannot be established with the dynamical process performing on the fly neighbor assignment introduced in Sec.~\ref{sec:random} . We therefore have to rely on some approximations. 

Consider the networks $\gamma_1$, $\gamma_2$, and $\gamma_b$ resulting from the three-networks decomposition presented in Sec.~\ref{sec:overlay}. Let $[X_1Y_2]_{ijk}(t)$ be the mean fraction of nodes in the system that are of 1-state $X_1$, of 2-state $Y_2$, have $i$ unmatched stubs on $\gamma_1$, $j$ unmatched stubs on $\gamma_2$, and $k$ unmatched stubs on $\gamma_b$ at time $t$. Once again, let us concentrate on the part of each ODE that corresponds to agent 1 dynamics.

In Sec.~\ref{sec:random}, the case of random overlap was considered in the asymptotic limit $N\to\infty$. Since one has $\Gamma_1=\gamma_1$ and $\Gamma_2=\gamma_2$ under this condition, substituting $[X_1 Y_2]_{ij}$ by $[X_1 Y_2]_{ijk}$ in Eqs. \eqref{eq:Theta1}--\eqref{eq:dRYdtNO} yields the correct description of agent 1 dynamics on $\gamma_1$. Hence, we only need to derive the additional contributions corresponding to the dynamics of agent 1 on $\gamma_b$. 

The approximation that we will use to take into account the overlap between $\Gamma_1$ and $\Gamma_2$ can be illustrated by the following example. Consider two nodes, node $n$ (of state $[X_1Y_2]_{ijk}$) and node $n'$ (of state $[X'_1Y'_2]_{i'j'k'}$), that are neighbors on $\Gamma_1$ and $\Gamma_2$, and that are just being involved together in an infectious contact with agent 1. From the point of view of a dynamical process performing on the fly neighbor assignment, their respective number $k$ and $k'$ of unmatched stubs on $\gamma_b$ should be decreased by one. However, $n$ and $n'$ could be later involved together in an agent 2 infectious contact. Since we do not track the information about the states of a node's neighbors in the formalism, one could account for this by increasing their number $j$ and $j'$ of unmatched stubs on $\gamma_2$ by one at the same moment $k$ and $k'$ are decreased. In other words, the prior information that $n$ and $n'$ were neighbors would be forgotten, but the fact that they may later be involved together in an agent 2 infectious contact is approximatively accounted for by granting them a new unmatched stub on $\gamma_2$. If an infectious contact occurs later over one of those stubs, it will be matched with another stub chosen at random between all the unmatched stubs on $\gamma_2$. 

Yet by doing so, we are throwing some useful information away. When two nodes are involved in an infectious contact, they share information about their respective states. If node $n$ was $[I_1S_2]_{ijk}$ and node $n'$ was $[S_1R_2]_{i'j'k'}$, the knowledge that node $n'$ is 2-recovered tells us that a transmission of agent 2 \emph{will never occur later between them}. Giving them an unmatched stub on $\gamma_2$ would thus be ill-advised; the  $ijk \to ij(k\!-\!1)$ and $i'j'k' \to i'j'(k'\!-\!1)$ transitions would be more appropriate. In summary, if node $n$ is involved in an agent 1 infectious contact via $\gamma_b$ with node $n'$, they will be given an additional unmatched stub on $\gamma_2$ if and only if their states are such that an agent 2 transmission may occur between them later in time. 

Let $\theta_{b,1}^Y$ be the probability that an unmatched stub on $\gamma_b$ belongs to a 1-infectious node of 2-state $Y_2$:
\begin{align}
\theta_{b,1}^Y = \dfrac{\sum_{i,j,k} k[I_1Y_2]_{ijk}}{\sum_{X'\!,Y'} \sum_{i'\!,j'\!,k'} k'[X'_1Y'_2]_{i'j'k'}} \ .
\label{eq:thetab1}
\end{align}
The total probability $\Theta_{b,1}$ that an unmatched stub on $\gamma_b$ belongs to a 1-infectious node is then given by $\Theta_{b,1}=\sum_Y \theta_{b,1}^Y$. We further define the probability $\Theta_{b,1}^Y$ that an unmatched stub on $\gamma_b$ belongs to a 1-infectious node \emph{whose 2-state is such that a transmission of agent 2 may occur later with a node of 2-state $Y_2$}: 
\begin{align}
\Theta_{b,1}^Y =
\begin{cases}
 \theta_{b,1}^S+\theta_{b,1}^I &  \textrm{if} \ Y\! =\! S \\
 \theta_{b,1}^S &  \textrm{if} \ Y\! =\! I \\
 0 &  \textrm{if} \ Y\! =\! R \ .
 \end{cases}
\end{align}
The probability of the opposite event is $\overline{\Theta}_{b,1}^Y\equiv \Theta_{b,1}\!-\! \Theta_{b,1}^Y$.

Similarly to Eq.~\eqref{eq:thetab1}, we define the probability $\phi_{b,1}^Y$ that any unmatched stub on $\gamma_b$ belongs to a node of particular 2-state $Y_2$:
\begin{align}
\phi_{b,1}^Y = \dfrac{\sum_X\sum_{i,j,k} k[X_1Y_2]_{ijk}}{\sum_{X'\!,Y'} \sum_{i'\!,j'\!,k'} k'[X'_1Y'_2]_{i'j'k'}} \ .
\end{align}
Naturally, $\sum_Y \phi_{b,1}^Y=1$. The probability $\Phi_{b,1}^Y$ that an unmatched stub on $\gamma_b$ belongs to any node whose 2-state is such that a transmission of agent 2 may occur later with a node of 2-state $Y_2$ is given by: 
\begin{align}
\Phi_{b,1}^Y =
\begin{cases}
 \phi_{b,1}^S+\phi_{b,1}^I &  \textrm{if} \ Y\! =\! S \\
 \phi_{b,1}^S &  \textrm{if} \ Y\! =\! I \\
 0 &  \textrm{if} \ Y\! =\! R \ .
 \end{cases}
\end{align}
We denote the complementary probability $\overline{\Phi}_{b,1}^Y\! \equiv \! 1- \Phi_{b,1}^Y$. 

First, a node in the $[S_1Y_2]_{ijk}$ compartment loses unmatched stubs on $\gamma_b$ at the rate $\beta_1 \Theta_{b,1} k$ as it suffers infectious contacts with 1-infectious individuals. This rate is composed of two distinct parts: infectious contacts from 1-infectious nodes with whom a transmission of agent 2 may occur later ($\beta_1 \Theta_{b,1}^Y k$) and from 1-infectious nodes with whom a transmission of agent 2 may never occur ($\beta_1 \overline{\Theta}_{b,1}^Y k$). As mentioned earlier, the first part yields a $ijk \to i(j\!+\!1)(k\!-\!1)$ index transition, while the second gives the transition $ijk\to ij(k\!-\!1)$. After an infectious contact, the $[S_1Y_2]_{ijk}$ node becomes 1-infectious with probability $\sigma_1^Y$, or remains 1-susceptible with probability $\overline{\sigma}_1^Y$. Combining those four different issues, the contribution of agent 1 dynamics on $\gamma_b$ to the ODE governing $[S_1Y_2]_{ijk}$ is
\begin{equation}
\begin{gathered}
\beta_1 (k\!+\!1) \overline{\sigma}_1^Y \! \left( \Theta_{b,1}^Y\! [S_1Y_2]_{i(j-1)(k+1)} + \overline{\Theta}_{b,1}^Y\! [S_1Y_2]_{ij(k+1)} \right) \\
- \beta_1 \Theta_{b,1} k [S_1Y_2]_{ijk} \ .
\label{eq:dSYdtO}
\end{gathered}
\end{equation}

Second, a node in the $[I_1Y_2]_{ijk}$ compartment loses unmatched stubs on $\gamma_b$ at the rate $\beta_1(1\!+\!\Theta_{b,1})k$ as it is the source or target of an infectious contact. Once again, this rate is composed of two parts that yield different index transitions: infectious contacts with a node with whom a transmission of agent 2 may \big[$\beta_1 (\Phi_{b,1}^Y\!+ \Theta_{b,1}^Y) k$\big] or may not \big[$\beta_1 (\overline{\Phi}_{b,1}^Y\!+ \overline{\Theta}_{b,1}^Y) k$\big] occur later. Taking into account the flows of incoming 1-susceptible nodes, the contribution of agent 1 dynamics on $\gamma_b$ to ODE for $[I_1Y_2]_{ijk}$ reads
\begin{equation}
\begin{gathered}
\beta_1 (k\! +\! 1)\sigma_1^Y \! \left( \Theta_{b,1}^Y \! [S_1Y_2]_{i(j-1)(k+1)} + \overline{\Theta}_{b,1}^Y \! [S_1Y_2]_{ij(k+1)} \right)  \\
+ \beta_1 (k\! +\! 1) \left( \Phi_{b,1}^Y\! [I_1Y_2]_{i(j-1)(k+1)} + \overline{\Phi}_{b,1}^Y\! [I_1Y_2]_{ij(k+1)} \right) \\
+ \beta_1 (k\! +\! 1) \left( \Theta_{b,1}^Y \! [I_1Y_2]_{i(j-1)(k+1)} + \overline{\Theta}_{b,1}^Y\! [I_1Y_2]_{ij(k+1)} \right) \\
 - \beta_1(1+\Theta_{b,1})k[I_1Y_2]_{ijk} \ .
\label{eq:dIYdtO}
\end{gathered}
\end{equation}
Finally, a node in the $[R_1Y_2]_{ijk}$ compartment loses unmatched stubs on $\gamma_b$ at the rate $\beta_1 \Theta_{b,1} k$ as it suffers infectious contacts with 1-infectious individuals. Taking into account both possible issues, i.e., if an unmatched stub is added afterwards on $\gamma_2$ or not, one obtains the contribution of agent 1 dynamics on $\gamma_b$ to the ODE governing $[R_1Y_2]_{ijk}$:
\begin{equation}
\begin{gathered}
\beta_1 (k\! +\! 1) \left( \Theta_{b,1}^Y\! [R_1Y_2]_{i(j-1)(k+1)} +\overline{\Theta}_{b,1}^Y \! [R_1Y_2]_{ij(k+1)} \right) \\
 - \beta_1\Theta_{b,1}k[R_1Y_2]_{ijk} \ .
\label{eq:dRYdtO}
\end{gathered}
\end{equation}
The interested reader is once again referred to the supplementary document for the complete system of ODEs~\footnotemark[1]. 

Initial conditions for agent 1 dynamics are given at $t=0$ by 
\begin{align}
[X_1Y_2]_{ijk} (0) = 
\begin{cases} 
(1\! -\! \epsilon_1)\rho(i,j,k) & \textrm{if} \ X\! =\! S \ \textrm{and} \ Y\! =\! S \\
\epsilon_1\rho(i,j,k) & \textrm{if} \ X\! =\! I \ \textrm{and} \ Y\! =\! S \\
0 & \textrm{otherwise} \ ,
\end{cases} 
\end{align}
while agent 2 dynamics is initialized at $t=\tau$ with
\begin{align}
[X_1Y_2]_{ijk} (\tau) \to
\begin{cases} 
(1\! -\! \epsilon_2)[X_1 S_2]_{ijk}(\tau) & \textrm{if} \ Y\! =\! S \\
\epsilon_2[X_1S_2]_{ijk}(\tau) & \textrm{if} \ Y\! =\! I \\
0 & \textrm{otherwise} \ .
\end{cases} 
\end{align}

The complexity of the ODE system derived in the case of random overlap increases as $\mathcal{O}(k_1^\textrm{max}\! \times \! k_2^\textrm{max})$, while it increases as $\mathcal{O}(\kappa_1^\textrm{max}\! \times \! \kappa_2^\textrm{max}\! \times \! \kappa_b^\textrm{max})$ in the case of arbitrary overlap. Here $\boldsymbol{k}^\textrm{max}$ and $\boldsymbol{\kappa}^\textrm{max}$ represent the largest degrees beyond which the systems of ODEs are truncated. Although high complexity may seem to be a major drawback of our approach, a significant speed up on Monte Carlo simulations can be obtained. 

\subsection{Other types of dynamics \label{sec:other}}

As we have pointed out at the end of Sec.~\ref{sec:SIR}, other types of interaction dynamics could have been considered. We give two examples to show how such alternative interaction rules could have been analytically handled using similar modeling schemes. 

First, suppose we allow for more than one infectious contact between individuals. In this case, compartmentalizing nodes by their number of unmatched stubs would be ill-advised. Indeed, when allowing for more than one infectious contact, the propagation dynamics cannot be made equivalent to a dynamical process performing on the fly neighbor assignment. Opting for a local description where nodes are sorted according to the number and state of their neighbors, as in~\cite{marceau10_pre,zombies}, would yield a more accurate description of the dynamics. Here, the basic state variables would read $[X_1Y_2]_{i_S i_I j_S j_I}(t)$, where $i_S$ and $i_I$ represent the number of neighbors on $\Gamma_1$ that are respectively 1-susceptible and 1-infectious (same for $j_S$ and $j_I$ regarding agent 2 on $\Gamma_2$). Note that tracking the number of 1- and 2-recovered neighbors is optional. 

Second, the leaky partial immunity rule could be replaced by perfect partial immunity~\cite{bansal09_arxiv}. In the latter, each $g$-susceptible node that is of $\hat{g}$-state $Y_{\hat{g}}$ regarding the other viral agent $\hat{g}$ has a probability $\overline{\sigma}_g^Y$ of being \emph{perfectly} immune to agent $g$. This scenario could be analytically modeled by introducing a fourth state compartment for each viral agent, $V_g$, denoting $g$-susceptible nodes that are perfectly immune to agent $g$.  

Finally, our analytical approach could also be made compatible with SIS dynamics after slight changes in the compartmentalization scheme. Because SIS dynamics cannot be made equivalent to a dynamical process performing on the fly neighbor assignment, it would once again be necessary to opt for a local description where nodes are sorted according to the number and state of their neighbors~\cite{marceau10_pre,zombies}.

\section{Validation through case studies \label{sec:results}}

In order to analyze the behavior of the model and to validate the accuracy of our mean-field approach, we consider two host populations, $A$ and $B$, which are under the threat of an agent 1 outbreak. Each host population is characterized by a different network $\Gamma_1$. For population $A$, the network $\Gamma_1$ is fairly homogeneous and features a Poisson degree distribution,
\begin{align}
P_1^A (k_1) = C_1^A \frac{\lambda_1^{k_1} e^{-\lambda_1}}{k_1!} \ , \ 0 \leq k_1 \leq 20 \ , \label{eq:p1a}
\end{align}
where $\lambda_1=3.5$ and $C_1^A$ is a normalization constant. We assume that the population $B$ displays more heterogeneous contact patterns, and has a network $\Gamma_1$ that follows a power-law degree distribution:
\begin{align}
P_1^B (k_1) = C_1^B k_1^{-\tau_1} \ , \ 1 \leq k_1 \leq 20 \ , \label{eq:p1b}
\end{align}
where $\tau_1=1.5$ and $C_1^B$ is a normalization constant. The parameters are chosen such that both networks have the same mean degree $\langle k_1 \rangle \simeq 3.5$ while featuring different level of heterogeneity. 

\begin{figure}[!t]
\includegraphics[width = .90\columnwidth]{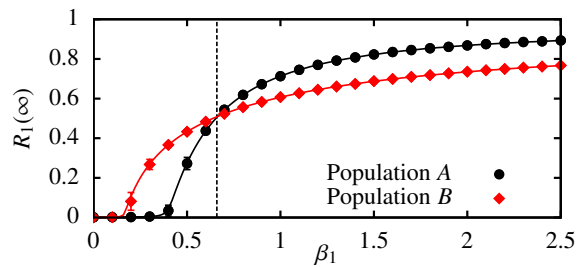} 
\caption{(Color online) Total agent 1 incidence $R_1(\infty)$ versus the infectious contact rate $\beta_1$ for a single viral agent SIR epidemic in two host populations with different network topologies. Population $A$ is characterized by a Poisson distributed network, while population $B$ has a power-law distributed network. Both networks feature the same mean degree $\langle k_1 \rangle \simeq 3.5$. Note that near $\beta_1=0.66$ (vertical line), both epidemics have a similar incidence. Points and error bars correspond to the mean and standard deviation of Monte Carlo simulations; solid curves are the predictions computed from our analytical approach. \label{fig:bif_z3p5_1d} }
\end{figure}

Figure~\ref{fig:bif_z3p5_1d} shows the phase diagram for the total \emph{incidence} of agent 1, defined by
\begin{align}
R_1(\infty) = \sum_Y \sum_{i,j,k} [R_1 Y_2]_{ijk} (\infty) \ ,
\end{align}
as a function of the infectious contact rate $\beta_1$ in the case where agent 1 propagates alone in the host populations $A$ and $B$. $R_1(\infty)$ is computed by setting $\epsilon_2=0$, in which case the dynamics reduces to the case of a single viral agent SIR epidemic. Note that for $\beta_1=0.66$, both epidemics reach approximatively the same level of incidence. For this reason, we will use this particular $\beta_1$ value throughout this section when comparing together both host populations. Moreover, Fig.~\ref{fig:bif_z3p5_1d} shows that our mean-field approach is able to capture with great accuracy the behavior of the model in the single-agent case. 

\subsection{Delayed intervention \label{sec:delay}}

In response to an epidemic menace, intervention strategies involving the propagation of a second viral agent in the host population may be employed to control the outbreak of the undesirable agent. Examples of such strategies include the spread of preventive information in the context of an emerging disease, or the dissemination of countermeasures to minimize the damages of a computer virus outbreak. In this section, we use the framework of interacting epidemics on overlay networks to analyze the efficiency of a hypothetical delayed agent 2 intervention on the outbreak of agent 1 in the host populations $A$ and $B$. 

 \begin{figure}[!t]
\subfigure[]{\includegraphics[width = .90\columnwidth]{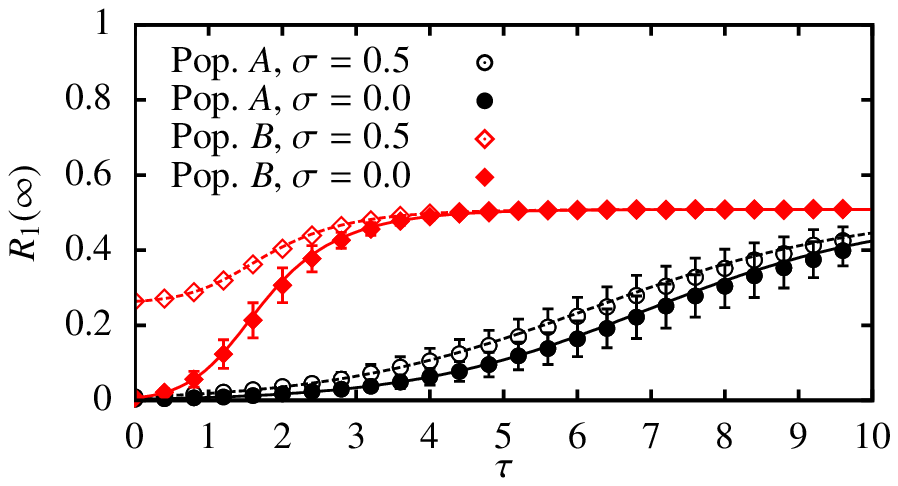} \label{fig:delay_z3p5}} \\ 
\mbox{
\subfigure[\ Population $A$]{\includegraphics[width = .5\columnwidth]{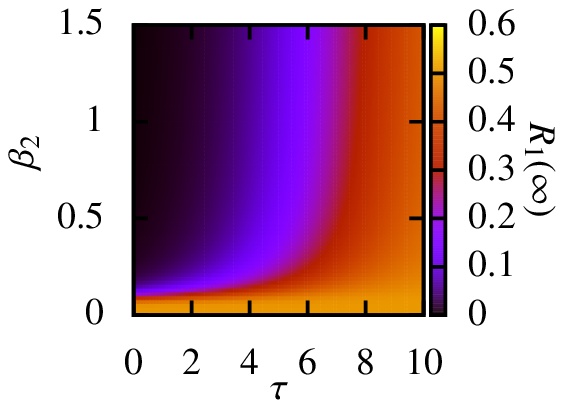} \label{fig:delay2d_z3p5_poisson}} 
\subfigure[\ Population $B$]{\includegraphics[width = .5\columnwidth]{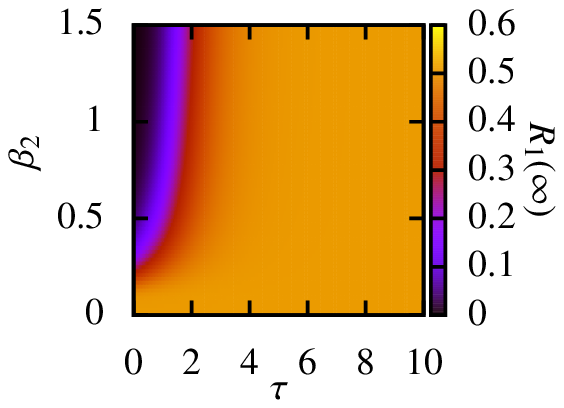} \label{fig:delay2d_z3p5_powerlaw}}} 
\caption{(Color online) (a) Total agent 1 incidence $R_1(\infty)$ in the host populations $A$ and $B$ versus the delay $\tau$ before an agent 2 intervention ($\beta_2=1$) providing full ($\sigma=0$) or partial ($\sigma=0.5$) immunity to agent 1. (b)-(c) Phase diagram showing $R_1(\infty)$ in both host populations as a function of the delay $\tau$ and the infectious contact rate $\beta_2$ of an agent 2 intervention providing full immunity to agent 1. $\beta_1=0.66$ in all figures. In both population, agent 2 spreads on a power-law distributed $\Gamma_2$ network that has a random overlap with the $\Gamma_1$ network. Points and error bars correspond to the mean and standard deviation of Monte Carlo simulations; curves are the predictions computed from our analytical approach. \label{fig:delay} }
\end{figure}

We consider a case of unidirectional immunity. We assume that agent 2 is not affected by agent 1, i.e., $\sigma_2^X=1$ for all $X \in \{ S,I,R\}$, while nodes that are either 2-infectious or 2-recovered benefit from a given level of immunity to agent 1, i.e., $\sigma_1^S=1$ and $\sigma_1^I=\sigma_1^R\equiv\sigma$. In both host populations, the network $\Gamma_2$, on which agent 2 propagates, is characterized by a power-law degree distribution, 
\begin{align}
P_2(k_2) = C_2 k_2^{-\tau_2} \ , \ 1 \leq k_2 \leq 40 \ ,
\end{align}
where $\tau_2=1$ and $C_2$ is a normalization constant. For now, we assume that the overlap between the networks $\Gamma_1$ and $\Gamma_2$ is random, and that there is no degree correlation between them, such that $P(\boldsymbol{k})=P_1(k_1)P_2(k_2)$.
 
\paragraph{Full immunity.} The total agent 1 incidence $R_1(\infty)$ in host populations $A$ and $B$ as a function of the delay $\tau$ between epidemics is illustrated on Fig.~\ref{fig:delay_z3p5} for the case of full ($\sigma=0$) and partial ($\sigma=0.5$) immunity. Let us consider the case of full immunity in the first place. For $\sigma=0$, we observe in Fig.~\ref{fig:delay_z3p5} that $R_1(\infty)$ increases much faster with $\tau$ in population $B$ than in population $A$. This means that one disposes of a much shorter time to react efficiently if the population in which agent 1 spreads features an heterogeneous structure, such as a power-law distributed network. Figures \ref{fig:delay2d_z3p5_poisson} and \ref{fig:delay2d_z3p5_powerlaw}, which also include the dependency of $R_1(\infty)$ on the infectious contact rate $\beta_2$, further confirm this observation.

\begin{figure}[!t]
\includegraphics[width = .90\columnwidth]{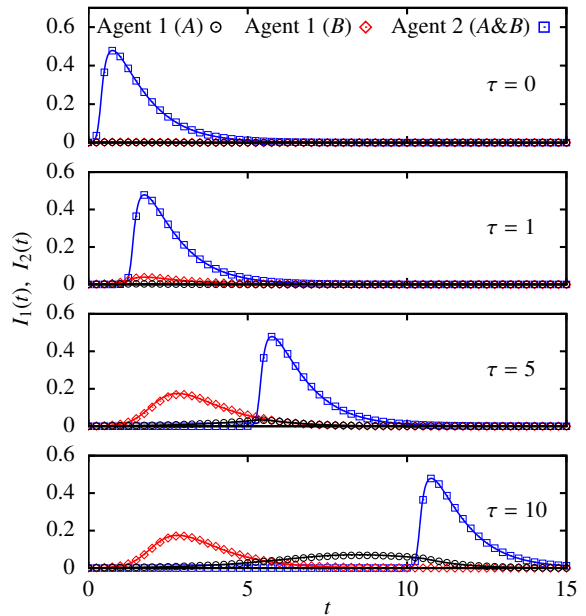} 
\caption{(Color online) Time evolution of the agent 1 and 2 prevalence, $I_1(t)$ and $I_2(t)$, in host populations $A$ and $B$ for various values of the delay $\tau$ before an agent 2 intervention providing full immunity ($\sigma=0$) to agent 1. Infectious contact rates are $\beta_1=0.66$ and $\beta_2=1$. Points corresponds to the mean of Monte Carlo simulations; error bars are of order of magnitude of the symbol size. Solid curves are the predictions computed from our analytical approach.  \label{fig:tevol_z3p5} }
\end{figure}

The results discussed in the last paragraph can be explained by looking at the time evolution of the epidemics. In Fig.~\ref{fig:tevol_z3p5}, the \emph{prevalence} at time $t$ of agent 1 and 2,
\begin{align}
I_1(t) = \sum_Y \sum_{i,j,k} [I_1 Y_2]_{ijk} (t)  \ , \ I_2(t) = \sum_X \sum_{i,j,k} [X_1 I_2]_{ijk}(t) \ ,
\end{align}
is illustrated for increasing values of the delay $\tau$ in the case of full immunity. For small values of $\tau$ ($\tau=0$ and $1$), agent 2 is able to inhibit the initial phase of the agent 1 epidemic in both host populations. At intermediate $\tau$ values ($\tau=5$), the agent 1 epidemic is still strongly restrained by agent 2 in population $A$, while it has almost enough time to run its course completely in  population $B$.  As $\tau$ is further increased ($\tau=10$), the effect of the intervention becomes minimal in both host populations. Figure~\ref{fig:tevol_z3p5} shows that the time scale of the agent 1 epidemic in a given host population is crucial in determining the efficiency of a delayed intervention. Since this time scale decreases with increasing network heterogeneity~\cite{barthelemy05_jtb}, this explains why much smaller values of $\tau$ are required in population $B$ to achieve an efficient intervention.

\paragraph{Partial immunity.} Let us now consider the case of partial immunity ($\sigma=0.5$). For the host population $A$, Fig.~\ref{fig:delay_z3p5} shows that the behavior of $R_1(\infty)$ versus $\tau$ is quite similar to the case of full immunity. The total agent 1 incidence increases slightly faster with the delay, which can be attributed to the fact that the fraction of the population reached by agent 2 before agent 1 is now partially vulnerable to the latter. However, the picture is drastically different for the host population $B$. Even when $\tau=0$, agent 2 is unable to inhibit the agent 1 epidemic, which reaches half as many nodes as it would reach without any intervention. 

This phenomenon is also a consequence of network heterogeneity. Consider a node of degree $k_1$ on $\Gamma_1$ that is still 1-susceptible and has acquired partial immunity to agent 1. If it is eventually involved in infectious contacts with all its neighbors on $\Gamma_1$, the probability that it remains 1-susceptible in the end is $(1-\sigma)^{k_1}$, which decreases exponentially with $k_1$. Therefore, leaky partial immunity has a weaker effect on high-degree nodes. This observation, together with the fact that the high-degree nodes are preferentially involved in the early phase of an epidemic~\cite{barthelemy05_jtb}, explains why agent 1 is able to invade a significant fraction of population $B$ even when a short-delay intervention is attempted. 

Finally, Figs.~\ref{fig:delay_z3p5} and \ref{fig:tevol_z3p5} show an excellent agreement between the analytical predictions computed from our mean-field approach and the outcome of Monte Carlo simulations of the dynamics. As we mentioned in Sec.~\ref{sec:random}, we expect our mean-field approach to be exact for configuration model overlay networks of infinite size and random overlap. The small divergence between the predictions of our approach and the mean values computed over Monte Carlo simulations can be attributed to finite-size effects, such as stochastic extinction at early times and the restriction on self-loops and repeated links when generating the networks.

\subsection{Overlap and degree correlation \label{sec:overlap}}

In the previous section, we assumed that the overlap between the networks $\Gamma_1$ and $\Gamma_2$ was random, and that there existed no degree correlation between them. We now relax this assumption and look at the effect of overlap and degree correlation. 

We consider a scenario very similar to that of the previous section, where agent 1 (disease) and agent 2 (intervention) propagate simultaneously in the host populations $A$ and $B$. We assume that the intervention is instantaneous ($\tau=0$) and grants full immunity to agent 1 ($\sigma=0$). The degree distributions of the network $\Gamma_1$ for population $A$ and $B$ are respectively given by \eqref{eq:p1a} and \eqref{eq:p1b}. Moreover, we now suppose in both populations that the degree distribution of $\Gamma_2$ is identical to that of $\Gamma_1$, i.e. $P_2(k)=P_1(k)\equiv p(k)$. 

\setcounter{paragraph}{0}
\paragraph{Overlap versus degree correlation.} In order to be able to isolate the respective effects of overlap and degree correlation, we distinguish between three different configurations: (i) random overlap and no degree correlation, in which case the system is built from $P(\boldsymbol{k})=p(k_1)p(k_2)$; (ii) random overlap and full degree correlation, in which case we build the system from $P(\boldsymbol{k})=p(k_1)\delta_{k_2,k_1}$; (iii) full overlap, in which case $\rho(\boldsymbol{\kappa})=\delta_{\kappa_1,0}\delta_{\kappa_2,0}p(\kappa_b)$ is used to generate the system. 

 \begin{figure}[!t]
\subfigure[\ Population $A$]{\includegraphics[width = .90\columnwidth]{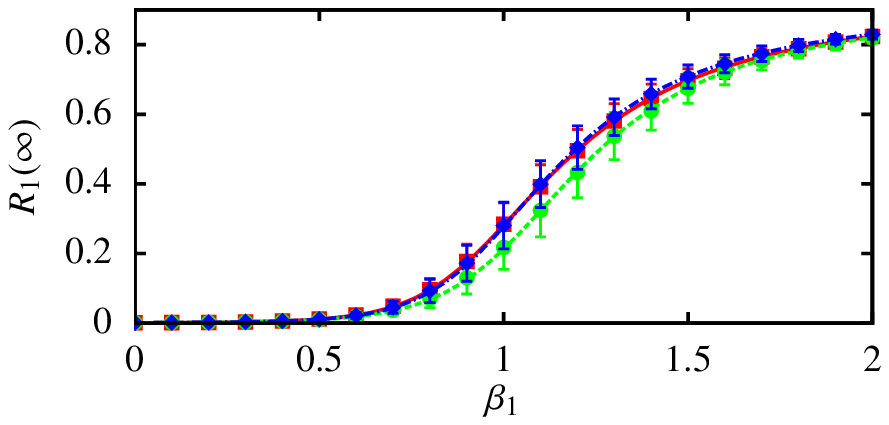} \label{fig:overlap_z3p5_poisson_full_1}} \\
\subfigure[\ Population $B$]{\includegraphics[width = .90\columnwidth]{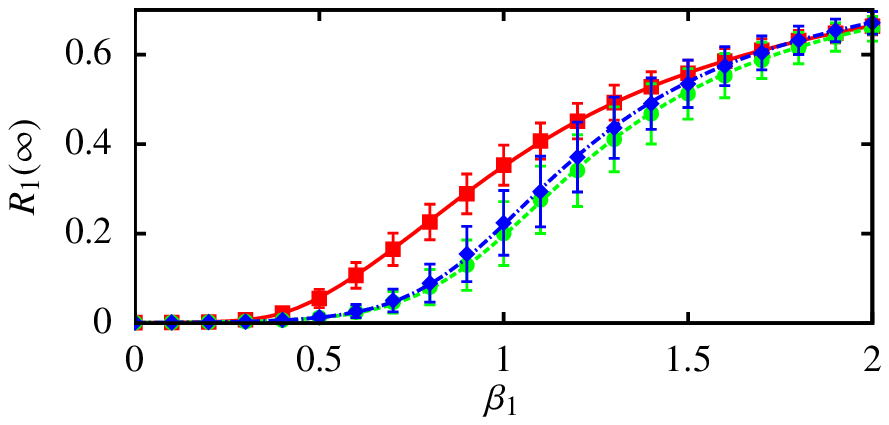} \label{fig:overlap_z3p5_powerlaw_full_1}}
\caption{(Color online) Total agent 1 incidence $R_1(\infty)$ versus the infectious contact rate $\beta_1$ in host populations $A$ and $B$ for different network configurations ({\color{Red}$\blacksquare$}: random overlap, no degree correlation; {\color{LimeGreen}\CIRCLE}: random overlap, full degree correlation; {\color{RoyalBlue}$\blacklozenge$}: full overlap). In all cases, the agent 2 intervention ($\beta_2=1$) is undelayed ($\tau=0$) and provides full immunity ($\sigma=0$) to agent 1. Points and error bars correspond to the mean and standard deviation of Monte Carlo simulations; curves are the predictions computed from our analytical approach. \label{fig:overlap_z3p5} }
\end{figure}

The bifurcation diagrams obtained for each host population using the three above configurations are illustrated in Fig.~\ref{fig:overlap_z3p5}. Let us compare in the first place the two configurations where the overlap is random. In both host populations, we observe that agent 1 manages to invade the system more easily in the case where there is no degree correlation between the networks. This is a consequence of the fact that the high-degree nodes of a network are more likely to be infected in an outbreak. If the high-degree nodes on $\Gamma_1$ and $\Gamma_2$ correspond to the same individuals, both viral agents will preferentially compete for their infection. In the scenario considered here, it will therefore be harder for agent 1 to invade the system if the high-degree nodes on $\Gamma_1$ are preferentially immunized by agent 2. The difference in heterogeneity between the host populations $A$ and $B$ explains why the effect of degree correlation is much stronger in $B$. 

Because full overlap implies full degree correlation, the fully overlapping configuration must be compared with the case of random overlap but full degree correlation in order to isolate the effect of overlap. In Fig.~\ref{fig:overlap_z3p5}, we observe that a configuration with full overlap facilitates the epidemic of agent 1 in the system. Since $\Gamma_1=\Gamma_2$ in the case of full overlap, nodes reached by agent 1 that are not already part of the agent 2 outbreak are more likely to be connected to other nodes that have not yet been reached by  agent 2. For a given joint probability distribution $P(\boldsymbol{k})$, this explains why invasion of agent 1 is easier in the fully overlapping case. 

Note that the magnitude of the difference between $R_1(\infty)$ in all the different configurations of degree correlation and overlap is larger when the time scales of the agent 1 and 2 epidemics are comparable. When $\beta_1$ is too low, agent 1 can barely invade the system in all configurations, while when $\beta_1 \gg \beta_2$, agent 1 is almost unaffected by agent 2 because the former spreads significantly faster. 

Our results corroborate the previous findings of Funk and Jansen. In~\cite{funk10_pre}, they considered the case where two processes, the first granting full immunity to the second, propagate subsequently on overlay networks. They showed that the epidemic threshold of the second process increases with positive degree correlation between networks with random overlap, while it decreases with increasing overlap for a given joint degree distribution $P(\boldsymbol{k})$. In our work, variations in the epidemic threshold are hard to analyze because both processes spread simultaneously in the host population. Our results generalizes the previous conclusions of~\cite{funk10_pre} to the case of dynamically interacting processes on overlay networks. 

\begin{figure}[!t]
\includegraphics[width = .90\columnwidth]{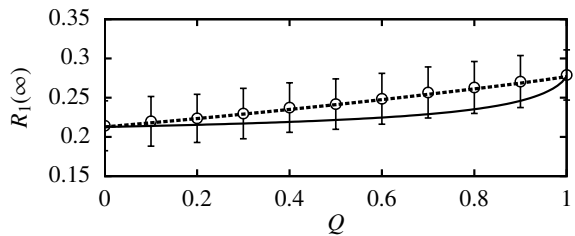} 
\caption{Total agent 1 incidence $R_1(\infty)$ in the host population $A$ versus the percentage of overlap $Q$ between $\Gamma_1$ and $\Gamma_2$. The parameters are $\sigma=0$, $\tau=0$, and $\beta_1=\beta_2=1$. Points and error bars correspond to the mean and standard deviation of 2500 Monte Carlo simulations performed with $N=100\, 000$. The solid curve corresponds to the predictions of the analytical approach developed in Sec.~\ref{sec:arbitrary}. The dashed curve corresponds to the analytical predictions obtained after making the substitutions given in Eqs.~\eqref{eq:subs}. \label{fig:partialoverlap} }
\end{figure}

\paragraph{Partial overlap.} In Sec.~\ref{sec:arbitrary}, some approximations were introduced in our mean-field approach to allow for arbitrary overlap between $\Gamma_1$ and $\Gamma_2$. To investigate their validity, we now consider the general case where the system is built from the following joint degree distribution
\begin{align}
\rho(\boldsymbol\kappa) = p(\kappa_1+\kappa_b) \delta_{\kappa_1,\kappa_2} \binom{\kappa_1+\kappa_b}{\kappa_b}Q^{\kappa_b}(1-Q)^{\kappa_1} \ ,
\end{align}
where $Q$ is the mean percentage of overlap in the system.

In Fig.~\ref{fig:partialoverlap}, we compare the outcome of Monte Carlo simulations with the analytical predictions of the approach developed in Sec.~\ref{sec:arbitrary} (solid curve) for increasing values of $Q$ in the host population $A$. We see that our approach becomes less accurate as $Q$ increases. This mainly comes from the fact that nodes can be granted additional unmatched stubs on $\gamma_1$ and $\gamma_2$ after a neighbor assigment on $\gamma_b$, which causes the dynamics on $\gamma_1$ and $\gamma_2$ to looses its \emph{exact} character. Figure \ref{fig:partialoverlap} however shows that our approach becomes accurate again as $Q$ gets very close to one. While this may appear counterintuitive at first, this is because an almost exact correspondence with the propagation dynamics is reobtained at $Q=1$. Since there are no links on the $\gamma_1$ and $\gamma_2$ networks at $Q=1$, the $i$ and $j$ indices are initially zero for each node. Thus, they are only used to indicate the number of new stubs that are granted after infectious contacts on $\gamma_b$, i.e., there is no mix between \emph{real} and \emph{supposed} unmatched stubs on $\gamma_1$ and $\gamma_2$.

Finally, it is interesting to note that for the special case of full immunity, an approach that handles partial overlap more accurately can be obtained. Indeed, since agent 2 provides full immunity to agent 1, we may use the following substitutions:
\begin{align}
\begin{aligned}
 \Theta_{b,2}^Y\to 0 \ \ &, \ \ \ \overline{\Theta}_{b,2}^Y\to \Theta_{b,2} \ , \\
 \Phi_{b,2}^Y\to 0 \ \ &, \ \ \ \overline{\Phi}_{b,2}^Y\to 1 \ ,
 \end{aligned} \label{eq:subs}
\end{align} 
for $Y \in \{ S,I,R \}$. While this modified approach yields a better description of the dynamics for the epidemic scenario considered here (dashed curve in Fig.~\ref{fig:partialoverlap}), we emphasize the fact that it is not valid for the general case of an interaction rule involving partial immunity.

\section{Conclusion \label{sec:conclusion}}

In this paper, we have introduced a general model where two viral agents propagate simultaneously on overlay networks and interact dynamically via a mechanism of leaky partial immunity. Exploiting a correspondence between propagation on networks and a dynamical process performing on the fly network generation~\cite{noel11_arxiv}, we have developed a network-based compartmental formalism in which nodes are sorted at any time according to their state and number of unmatched stubs. Unlike previous work based on bond percolation~\cite{newman05_prl,bansal09_arxiv,bansal10_plos,funk10_pre}, our analytical approach gives the complete time evolution of the system. By direct comparison with full Monte Carlo simulations of the model, we have demonstrated that it is able to capture with great accuracy the dynamical interaction between epidemics occurring simultaneously on overlay networks featuring various level of heterogeneity, degree correlation, and overlap. 

Our analytical approach is highly versatile and may be applied to numerous scenarios of diverse nature. Here, we have considered a hypothetical delayed intervention scenario, in which an immunizing process is disseminated in a host population to hinder the propagation of an undesirable process. We have discussed the influence of the delay and the level of immunity on the intervention efficiency in host populations featuring homogeneous and heterogeneous network structures. Moreover, we have shown that positive degree correlation increases the efficiency of the intervention, while overlap facilitates the invasion of the undesirable process. 

Finally, this work highlights the power of the general modeling scheme presented in~\cite{noel11_arxiv}, from which our formalism stems. Part of ongoing research focuses on the application of these guidelines to include more realistic features in our model, such as community structured networks~\cite{hebert10_pre}. Other directions for future research include the application of our analytical approach to investigate specific case studies, such as the influence of other sexually transmitted infections on the spread of HIV~\cite{boily09_lancet}. 

\begin{acknowledgments}
Our research team is grateful to FQRNT (V.M., L.H.-D., and L.J.D.), NSERC (V.M. and L.J.D.), and CIHR (P.-A.N., L.H.-D., and A.A.) for financial support.
\end{acknowledgments}


\onecolumngrid
\renewcommand{\thesection}{S-\Roman{section}}
\newdimen\oldparindent\oldparindent=\parindent
\setcounter{section}{0}

\newpage 

\begin{center}
{\large \bf SUPPLEMENTARY DOCUMENT \\
for \\
``Modeling the dynamical interaction between epidemics on overlay networks''}
\end{center}

\begin{center}
Vincent Marceau, Pierre-Andr\'e No\"el, Laurent H\'ebert-Dufresne, Antoine Allard, and Louis J. Dub\'e \\
{\small \it D\'epartement de Physique, de G\'enie Physique, et d'Optique, Universit\'e Laval, Qu\'ebec, Qu\'ebec, Canada G1V 0A6}
\end{center}

\begin{center}
\begin{minipage}{0.77\textwidth}{\small \parindent=\oldparindent
This supplementary document contains the full set of equations of the analytical approach presented in Marceau \emph{et al.}, ``Modeling the dynamical interaction between epidemics on overlay networks.'' The notation used is presented in the first place in Sec.~\ref{sec:SDnotation}. Equations for the case of overlay networks with random overlap are presented in Sec.~\ref{sec:SDrandom}, and the equations for overlay networks with arbitrary overlap appear in Sec.~\ref{sec:SDarbitrary}.}
\end{minipage}
\end{center}

\section{Notation \label{sec:SDnotation}}

\setlength{\extrarowheight}{3pt}
\begin{table}[!h]
\caption{Notation used in this document}
\begin{center}
\begin{tabular}{@{\extracolsep{2pt}}|c|m{0.75\textwidth}|}
\hline
$P(k_1,k_2)$ & Joint degree distribution of the system of overlay networks consisting of $\Gamma_1$ and $\Gamma_2$. \\ \hline
$\rho(\kappa_1,\kappa_2,\kappa_b)$ & Joint degree distribution of the three non-overlapping networks decomposition consisting of $\gamma_1$, $\gamma_2$, and $\gamma_b$.  \\ \hline
$[X_1 Y_2]_{ij}(t)$ & Mean fraction of nodes in the host population that are of 1-state $X_1$, of 2-state $Y_2$, have $i$ unmatched stubs on $\Gamma_1$, and $j$ unmatched stubs on $\Gamma_2$ at time $t$. \\ \hline 
$[X_1 Y_2]_{ijk}(t)$ & Mean fraction of nodes in the host population that are of 1-state $X_1$, of 2-state $Y_2$, have $i$ unmatched stubs on $\gamma_1$, $j$ unmatched stubs on $\gamma_2$, and $k$ unmatched stubs on $\gamma_b$ at time $t$. \\ \hline 
$g$\ , \ $\hat{g}$ & The index $g\in\{1,2\}$ denotes the viral agent currently under consideration. The other viral agent is denoted by the index $\hat{g}\in \{2,1\}$. \\ \hline
$\alpha_g$ & Recovery rate for viral agent $g$. \\ \hline
$\beta_g$ & Infectious contact rate for viral agent $g$. \\ \hline
$\epsilon_g$ & Fraction of nodes in the host population that are initially $g$-infectious. \\ \hline
$\sigma_g^Y$ & Probability of successful transmission of viral agent $g$ after an infectious contact involving a $g$-susceptible node of $\hat{g}$-state $Y_{\hat{g}}$ regarding the other viral agent $\hat{g}$. \\ \hline
$\overline{\sigma}_g^Y$ & Probability of failed transmission of viral agent $g$ after an infectious contact involving a $g$-susceptible node of $\hat{g}$-state $Y_{\hat{g}}$ regarding the other viral agent $\hat{g}$ ($\overline{\sigma}_g^Y \equiv 1- \sigma_g^Y$). \\ \hline
$\Theta_g$ & Probability that an unmatched stub chosen at random on network $\gamma_g$ belongs to a $g$-infectious node. \\ \hline
$\Theta_{b,g}$ & Probability that an unmatched stub chosen at random on network $\gamma_b$ belongs to a $g$-infectious node. \\ \hline
$\Theta_{b,g}^Y$ & Probability that an unmatched stub chosen at random on network $\gamma_b$ belongs to a $g$-infectious node whose $\hat{g}$-state regarding the other viral agent $\hat{g}$ is such that a transmission of agent $\hat{g}$ may occur later between it and a node of $\hat{g}$-state $Y_{\hat{g}}$. \\ \hline
$\overline{\Theta}_{b,g}^Y$ & Probability that an unmatched stub chosen at random on network $\gamma_b$ belongs to a $g$-infectious node whose $\hat{g}$-state regarding the other viral agent $\hat{g}$ is such that a transmission of agent $\hat{g}$ may never occur later between it and a node of $\hat{g}$-state $Y_{\hat{g}}$ ($\overline{\Theta}_{b,g}^Y \equiv \Theta_{b,g} - \Theta_{b,g}^Y$ ). \\ \hline
$\Phi_{b,g}^Y$ & Probability that an unmatched stub chosen at random on network $\gamma_b$ belongs to a node whose $\hat{g}$-state regarding the other viral agent $\hat{g}$ is such that a transmission of agent $\hat{g}$ may occur later between it and a node of $\hat{g}$-state $Y_{\hat{g}}$. \\ \hline
$\overline{\Phi}_{b,g}^Y$ & Probability that an unmatched stub chosen at random on network $\gamma_b$ belongs to a node whose $\hat{g}$-state regarding the other viral agent $\hat{g}$ is such that a transmission of agent $\hat{g}$ may never occur later between it and a node of $\hat{g}$-state $Y_{\hat{g}}$ ($\overline{\Phi}_{b,g}^Y \equiv 1 - \Phi_{b,g}^Y$ ). \\ \hline
\end{tabular}
\end{center}
\label{tab:notation}
\end{table}
\setlength{\extrarowheight}{0pt}

The following summation convention is used throughout this document:
\begin{align}
\sum_X \ \ \equiv \!\! \sum_{X \in \{S,I,R\} } \ , \qquad \sum_Y \ \ \equiv \!\! \sum_{Y \in \{S,I,R\} } \ , \qquad 
\sum_i \equiv  \sum_{i=0}^\infty \ , \qquad \sum_j \equiv  \sum_{j=0}^\infty \ , \qquad \sum_k \equiv  \sum_{k=0}^\infty \ .
\end{align}

\newpage

\section{Overlay networks with random overlap \label{sec:SDrandom}}

Let 
\begin{align}
\Theta_1 \equiv \dfrac{\sum_Y \sum_{i,j} i[I_1Y_2]_{ij}}{\sum_{X' ,Y'} \sum_{i' ,j'} i'[X'_1Y'_2]_{i'j'}}  \ \ \ \ \textrm{and} \ \ \ \
\Theta_2 \equiv \dfrac{\sum_X \sum_{i,j} j[X_1I_2]_{ij}}{\sum_{X' ,Y'} \sum_{i' ,j'} j'[X'_1Y'_2]_{i'j'}} \ .
\end{align}
For the case of overlay networks with random overlap, the following system of ordinary differential equations (ODEs) governs the time evolution of the $[X_1 Y_2]_{ij}$ compartments:
\begin{align}
&\begin{aligned}
\frac{d}{dt}[S_1 S_2]_{ij} = & \ \beta_1 \Theta_1   \left[ (i  +  1)\overline{\sigma}_1^S   [S_1 S_2]_{(i+1)j}   - i[S_1 S_2]_{ij} \right] \\ 
& + \beta_2 \Theta_2   \left[ (j  +  1)\overline{\sigma}_2^S   [S_1 S_2]_{i(j+1)}   - j[S_1 S_2]_{ij} \right] 
\end{aligned}\\
&\begin{aligned}
\frac{d}{dt}[S_1 I_2]_{ij} = & \ \beta_1 \Theta_1   \left[ (i  +  1)\overline{\sigma}_1^I   [S_1 I_2]_{(i+1)j}   - i[S_1 I_2]_{ij} \right] \\
& - \alpha_2 [S_1 I_2]_{ij} + \beta_2 \Theta_2 (j  +  1)\sigma_2^S [S_1 S_2]_{i(j+1)} + \beta_2(1  +   \Theta_2)   \left[ (j  +  1)[S_1I_2]_{i(j+1)}   - j[S_1 I_2]_{ij} \right] 
\end{aligned}\\
&\begin{aligned}
\frac{d}{dt}[S_1 R_2]_{ij} = & \ \beta_1 \Theta_1   \left[ (i  +  1)\overline{\sigma}_1^R [S_1 R_2]_{(i+1)j}   - i[S_1 R_2]_{ij} \right] \\
& + \alpha_2 [S_1 I_2]_{ij} + \beta_2 \Theta_2 \left[ (j  +  1)[S_1R_2]_{i(j+1)}   - j[S_1 R_2]_{ij} \right] 
\end{aligned}\\
&\begin{aligned}
\frac{d}{dt}[I_1 S_2]_{ij} = &  - \alpha_1 [I_1 S_2]_{ij} + \beta_1 \Theta_1 (i  +  1)\sigma_1^S [S_1 S_2]_{(i+1)j} + \beta_1(1  +   \Theta_1)   \left[ (i  +  1)[I_1S_2]_{(i+1)j}   - i[I_1 S_2]_{ij} \right] \\
& + \beta_2 \Theta_2   \left[ (j  +  1)\overline{\sigma}_2^I   [I_1 S_2]_{i(j+1)}   - j[I_1 S_2]_{ij} \right]
\end{aligned}\\
&\begin{aligned}
\frac{d}{dt}[I_1 I_2]_{ij} = &  - \alpha_1 [I_1 I_2]_{ij} + \beta_1 \Theta_1 (i  +  1)\sigma_1^I [S_1 I_2]_{(i+1)j} + \beta_1(1  +   \Theta_1)   \left[ (i  +  1)[I_1I_2]_{(i+1)j}   - i[I_1 I_2]_{ij} \right] \\ 
& - \alpha_2 [I_1 I_2]_{ij} + \beta_2 \Theta_2 (j  +  1)\sigma_2^I [I_1 S_2]_{i(j+1)} + \beta_2(1  +   \Theta_2)   \left[ (j  +  1)[I_1I_2]_{i(j+1)}   - j[I_1 I_2]_{ij} \right] 
\end{aligned}\\
&\begin{aligned}
\frac{d}{dt}[I_1 R_2]_{ij} = &  - \alpha_1 [I_1 R_2]_{ij} + \beta_1 \Theta_1 (i  +  1)\sigma_1^R [S_1 R_2]_{(i+1)j} + \beta_1(1  +   \Theta_1)   \left[ (i  +  1)[I_1R_2]_{(i+1)j}   - i[I_1 R_2]_{ij} \right] \\ 
& + \alpha_2 [I_1 I_2]_{ij} + \beta_2 \Theta_2 \left[ (j  +  1)[I_1R_2]_{i(j+1)}   - j[I_1 R_2]_{ij} \right] 
\end{aligned}\\
&\begin{aligned}
\frac{d}{dt}[R_1 S_2]_{ij} = & \ \alpha_1 [I_1 S_2]_{ij} + \beta_1 \Theta_1 \left[ (i  +  1)[R_1S_2]_{(i+1)j}   - i[R_1 S_2]_{ij} \right] \\
& + \beta_2 \Theta_2   \left[ (j  +  1)\overline{\sigma}_2^R   [R_1 S_2]_{i(j+1)}   - j[R_1 S_2]_{ij} \right]
\end{aligned}\\
&\begin{aligned}
\frac{d}{dt}[R_1 I_2]_{ij} = &  \ \alpha_1 [I_1 I_2]_{ij} +  \beta_1 \Theta_1 \left[ (i  +  1)[R_1I_2]_{(i+1)j}   - i[R_1 I_2]_{ij} \right] \\ 
& - \alpha_2 [R_1 I_2]_{ij} + \beta_2 \Theta_2 (j  +  1)\sigma_2^R [R_1 S_2]_{i(j+1)} + \beta_2(1  +   \Theta_2)   \left[ (j  +  1)[R_1I_2]_{i(j+1)}   - j[R_1 I_2]_{ij} \right] 
\end{aligned}\\
&\begin{aligned}
\frac{d}{dt}[R_1 R_2]_{ij} = & \ \alpha_1 [I_1 R_2]_{ij} + \beta_1 \Theta_1 \left[ (i  +  1)[R_1R_2]_{(i+1)j}   - i[R_1 R_2]_{ij} \right] \\ 
& + \alpha_2 [R_1 I_2]_{ij} + \beta_2 \Theta_2 \left[ (j  +  1)[R_1R_2]_{i(j+1)}   - j[R_1 R_2]_{ij} \right]  \ .
\end{aligned}
\end{align}
This system of ODEs satisfy the conservation of nodes, namely 
\begin{align}
\sum_{X, Y} \sum_{i,j} \frac{d}{dt} [X_1 Y_2]_{ij} = 0 \ .
\end{align}
The dynamics of agent 1 is initialized at time $t=0$ with the following initial conditions:
\begin{equation}
\begin{array}{@{\extracolsep{40pt}} l l l }
\ [S_1 S_2]_{ij}(0) = (1-\epsilon_1) P(i,j)  & [I_1 S_2]_{ij}(0) = \epsilon_1 P(i,j)  & [R_1 S_2]_{ij}(0) = 0 \\
\ [S_1 I_2]_{ij}(0) = 0  & [I_1 I_2]_{ij}(0) = 0  & [R_1 I_2]_{ij}(0) = 0 \\
\ [S_1 R_2]_{ij}(0) = 0  & [I_1 R_2]_{ij}(0) = 0  & [R_1 R_2]_{ij}(0) = 0 \ .
\end{array}
\end{equation}
The dynamics of agent 2 is initialized at time $t=\tau \geq 0$ by making the following substitutions:
\begin{equation}
\begin{array}{@{\extracolsep{15pt}} l l l }
\ [S_1 S_2]_{ij}(\tau) \to (1-\epsilon_2) [S_1 S_2]_{ij}(\tau)  & [I_1 S_2]_{ij}(\tau) \to (1-\epsilon_2) [I_1 S_2]_{ij}(\tau)  & [R_1 S_2]_{ij}(\tau) \to (1-\epsilon_2) [R_1 S_2]_{ij}(\tau) \\
\ [S_1 I_2]_{ij}(\tau) \to \epsilon_2 [S_1 S_2]_{ij}(\tau)   & [I_1 I_2]_{ij}(\tau) \to \epsilon_2 [I_1 S_2]_{ij}(\tau)   & [R_1 I_2]_{ij}(\tau) \to \epsilon_2 [R_1 S_2]_{ij}(\tau)  \ .
\end{array}
\end{equation}

\newpage
\section{Overlay networks with arbitrary overlap \label{sec:SDarbitrary}}

Let 
\begin{align}
\begin{aligned}
&\Theta_1 \equiv \dfrac{\sum_Y \sum_{i,j,k} i[I_1Y_2]_{ijk}}{\sum_{X' ,Y'} \sum_{i' ,j' ,k'} i'[X'_1Y'_2]_{i'j'k'}} \ \ , \ \ \ \
\Theta_2 \equiv \dfrac{\sum_X \sum_{i,j,k} j[X_1I_2]_{ijk}}{\sum_{X' ,Y'} \sum_{i' ,j' ,k'} j'[X'_1Y'_2]_{i'j'k'}} \ \ , \\
&\Theta_{b,1} \equiv \dfrac{\sum_Y \sum_{i,j,k} k[I_1Y_2]_{ijk}}{\sum_{X' ,Y'} \sum_{i' ,j' ,k'} k'[X'_1Y'_2]_{i'j'k'}} \ \ , \ \ \textrm{and} \ \
\Theta_{b,2} \equiv \dfrac{\sum_X \sum_{i,j,k} k[X_1I_2]_{ijk}}{\sum_{X' ,Y'} \sum_{i' ,j' ,k'} k'[X'_1Y'_2]_{i'j'k'}} \ .
\end{aligned}
\end{align}
We further define the following probabilities:
\begin{align}
&\begin{aligned}
\Theta_{b,1}^Y \equiv
\setlength{\extrarowheight}{7pt}
\begin{cases}
\dfrac{ \sum_{i,j,k} \left( k[I_1S_2]_{ijk} +  k[I_1I_2]_{ijk}\right)}{\sum_{X' ,Y'} \sum_{i' ,j' ,k'} k'[X'_1Y'_2]_{i'j'k'}}  &  \textrm{if} \ Y  =  S \\
\dfrac{ \sum_{i,j,k} k[I_1 S_2]_{ijk}}{\sum_{X' ,Y'} \sum_{i' ,j' ,k'} k'[X'_1Y'_2]_{i'j'k'}} &  \textrm{if} \ Y  =  I \\
 0 &  \textrm{if} \ Y  =  R 
 \end{cases}
 \setlength{\extrarowheight}{3pt}
 \ \ \ , \ \ \ \ \ \
 \overline{\Theta}_{b,1}^Y \equiv
\begin{cases}
 \Theta_{b,1} - \Theta_{b,1}^S &  \textrm{if} \ Y  =  S \\
 \Theta_{b,1} - \Theta_{b,1}^I  &  \textrm{if} \ Y  =  I \\
 \Theta_{b,1} - \Theta_{b,1}^R &  \textrm{if} \ Y  =  R 
 \end{cases}
 \ \ \ ,
\end{aligned} 
\end{align}
\begin{align}
&\begin{aligned}
\Phi_{b,1}^Y \equiv
\setlength{\extrarowheight}{7pt}
\begin{cases}
\dfrac{ \sum_X \sum_{i,j,k} \left( k[X_1S_2]_{ijk} +  k[X_1I_2]_{ijk}\right)}{\sum_{X' ,Y'} \sum_{i' ,j' ,k'} k'[X'_1Y'_2]_{i'j'k'}}  &  \textrm{if} \ Y  =  S \\
\dfrac{ \sum_X \sum_{i,j,k} k[X_1 S_2]_{ijk}}{\sum_{X' ,Y'} \sum_{i' ,j' ,k'} k'[X'_1Y'_2]_{i'j'k'}} &  \textrm{if} \ Y  =  I \\
 0 &  \textrm{if} \ Y  =  R 
 \end{cases}
 \setlength{\extrarowheight}{3pt}
 \ \ \ , \ \ \textrm{and} \ \
 \overline{\Phi}_{b,1}^Y \equiv
\begin{cases}
1 - \Phi_{b,1}^S &  \textrm{if} \ Y  =  S \\
1 - \Phi_{b,1}^I  &  \textrm{if} \ Y  =  I \\
1 - \Phi_{b,1}^R &  \textrm{if} \ Y  =  R 
 \end{cases}
 \ \ \ .
\end{aligned} 
\end{align}
\begin{align}
&\begin{aligned}
\Theta_{b,2}^Y \equiv
\setlength{\extrarowheight}{7pt}
\begin{cases}
\dfrac{ \sum_{i,j,k} \left( k[S_1I_2]_{ijk} +  k[I_1I_2]_{ijk}\right)}{\sum_{X' ,Y'} \sum_{i' ,j' ,k'} k'[X'_1Y'_2]_{i'j'k'}}  &  \textrm{if} \ Y  =  S \\
\dfrac{ \sum_{i,j,k} k[S_1 I_2]_{ijk}}{\sum_{X' ,Y'} \sum_{i' ,j' ,k'} k'[X'_1Y'_2]_{i'j'k'}} &  \textrm{if} \ Y  =  I \\
 0 &  \textrm{if} \ Y  =  R 
 \end{cases}
 \setlength{\extrarowheight}{3pt}
 \ \ \ , \ \ \ \ \ \
 \overline{\Theta}_{b,2}^Y \equiv
\begin{cases}
 \Theta_{b,2} - \Theta_{b,2}^S &  \textrm{if} \ Y  =  S \\
 \Theta_{b,2} - \Theta_{b,2}^I  &  \textrm{if} \ Y  =  I \\
 \Theta_{b,2} - \Theta_{b,2}^R &  \textrm{if} \ Y  =  R 
 \end{cases}
 \ \ \ ,
\end{aligned} 
\end{align}
\begin{align}
&\begin{aligned}
\Phi_{b,2}^Y \equiv
\setlength{\extrarowheight}{7pt}
\begin{cases}
\dfrac{ \sum_X \sum_{i,j,k} \left( k[S_1X_2]_{ijk} +  k[I_1X_2]_{ijk}\right)}{\sum_{X' ,Y'} \sum_{i' ,j' ,k'} k'[X'_1Y'_2]_{i'j'k'}}  &  \textrm{if} \ Y  =  S \\
\dfrac{ \sum_X \sum_{i,j,k} k[S_1 X_2]_{ijk}}{\sum_{X' ,Y'} \sum_{i' ,j' ,k'} k'[X'_1Y'_2]_{i'j'k'}} &  \textrm{if} \ Y  =  I \\
 0 &  \textrm{if} \ Y  =  R 
 \end{cases}
 \setlength{\extrarowheight}{3pt}
 \ \ \ , \ \ \textrm{and} \ \
 \overline{\Phi}_{b,2}^Y \equiv
\begin{cases}
1 - \Phi_{b,2}^S &  \textrm{if} \ Y  =  S \\
1 - \Phi_{b,2}^I  &  \textrm{if} \ Y  =  I \\
1 - \Phi_{b,2}^R &  \textrm{if} \ Y  =  R 
 \end{cases}
 \ \ \ .
\end{aligned} 
\end{align}
\setlength{\extrarowheight}{0pt}
For overlay networks with arbitrary overlap, the following ODEs govern the time evolution of the $[X_1 Y_2]_{ijk}$ compartments:
\begin{align}
&\begin{aligned}
\frac{d}{dt}[S_1 S_2]_{ijk} = & \ \beta_1 \Theta_1   \left[ (i  +  1)\overline{\sigma}_1^S   [S_1 S_2]_{(i+1)jk}   - i[S_1 S_2]_{ijk} \right] \\ & + \beta_1 (k + 1) \overline{\sigma}_1^S   \left( \Theta_{b,1}^S  [S_1S_2]_{i(j-1)(k+1)} + \overline{\Theta}_{b,1}^S  [S_1S_2]_{ij(k+1)} \right) - \beta_1 \Theta_{b,1} k [S_1S_2]_{ijk} \\ 
& + \beta_2 \Theta_2   \left[ (j  +  1)\overline{\sigma}_2^S   [S_1 S_2]_{i(j+1)k}   - j[S_1 S_2]_{ijk} \right] \\
& + \beta_2 (k + 1) \overline{\sigma}_2^S   \left( \Theta_{b,2}^S  [S_1S_2]_{(i-1)j(k+1)} + \overline{\Theta}_{b,2}^S  [S_1S_2]_{ij(k+1)} \right) - \beta_2 \Theta_{b,2} k [S_1S_2]_{ijk}
\end{aligned}\\
&\begin{aligned}
\frac{d}{dt}[S_1 I_2]_{ijk} = & \ \beta_1 \Theta_1   \left[ (i  +  1)\overline{\sigma}_1^I   [S_1 I_2]_{(i+1)jk}   - i[S_1 I_2]_{ijk} \right] \\
& + \beta_1 (k + 1) \overline{\sigma}_1^I   \left( \Theta_{b,1}^I  [S_1I_2]_{i(j-1)(k+1)} + \overline{\Theta}_{b,1}^I  [S_1I_2]_{ij(k+1)} \right) - \beta_1 \Theta_{b,1} k [S_1I_2]_{ijk} \\ 
& - \alpha_2 [S_1 I_2]_{ijk} + \beta_2 \Theta_2 (j  +  1)\sigma_2^S [S_1 S_2]_{i(j+1)k} + \beta_2(1  +   \Theta_2)   \left[ (j  +  1)[S_1I_2]_{i(j+1)k}   - j[S_1 I_2]_{ijk} \right] \\
& + \beta_2 (k  +  1)\sigma_2^S   \left( \Theta_{b,2}^S [S_1S_2]_{(i-1)j(k+1)} + \overline{\Theta}_{b,2}^S [S_1S_2]_{ij(k+1)} \right) \\ 
& + \beta_2 (k  +  1) \left[ \Big( \Phi_{b,2}^S   +   \Theta_{b,2}^S\Big) [S_1I_2]_{(i-1)j(k+1)} + \Big( \overline{\Phi}_{b,2}^S  +   \overline{\Theta}_{b,2}^S \Big) [S_1I_2]_{ij(k+1)} \right]  - \beta_2(1+\Theta_{b,2})k[S_1I_2]_{ijk} 
\end{aligned}\\
&\begin{aligned}
\frac{d}{dt}[S_1 R_2]_{ijk} = & \ \beta_1 \Theta_1   \left[ (i  +  1)\overline{\sigma}_1^R [S_1 R_2]_{(i+1)jk}   - i[S_1 R_2]_{ijk} \right] \\
& + \beta_1 (k + 1) \overline{\sigma}_1^R   \left( \Theta_{b,1}^R  [S_1R_2]_{i(j-1)(k+1)} + \overline{\Theta}_{b,1}^R  [S_1R_2]_{ij(k+1)} \right) - \beta_1 \Theta_{b,1} k [S_1R_2]_{ijk} \\ 
& + \alpha_2 [S_1 I_2]_{ijk} + \beta_2 \Theta_2 \left[ (j  +  1)[S_1R_2]_{i(j+1)k}   - j[S_1 R_2]_{ijk} \right] \\
& + \beta_2 (k  +  1) \left( \Theta_{b,2}^S  [S_1R_2]_{(i-1)j(k+1)} +\overline{\Theta}_{b,2}^S   [S_1R_2]_{ij(k+1)} \right)  - \beta_2\Theta_{b,2}k[S_1R_2]_{ijk}
\end{aligned}\\
&\begin{aligned}
\frac{d}{dt}[I_1 S_2]_{ijk} = &  - \alpha_1 [I_1 S_2]_{ijk} + \beta_1 \Theta_1 (i  +  1)\sigma_1^S [S_1 S_2]_{(i+1)jk} + \beta_1(1  +   \Theta_1)   \left[ (i  +  1)[I_1S_2]_{(i+1)jk}   - i[I_1 S_2]_{ijk} \right] \\
& + \beta_1 (k  +  1)\sigma_1^S   \left( \Theta_{b,1}^S [S_1S_2]_{i(j-1)(k+1)} + \overline{\Theta}_{b,1}^S [S_1S_2]_{ij(k+1)} \right) \\ 
& + \beta_1 (k  +  1) \left[ \Big( \Phi_{b,1}^S   +   \Theta_{b,1}^S\Big) [I_1S_2]_{i(j-1)(k+1)} + \Big( \overline{\Phi}_{b,1}^S  +   \overline{\Theta}_{b,1}^S \Big) [I_1S_2]_{ij(k+1)} \right]  - \beta_1(1+\Theta_{b,1})k[I_1S_2]_{ijk} \\
& + \beta_2 \Theta_2   \left[ (j  +  1)\overline{\sigma}_2^I   [I_1 S_2]_{i(j+1)k}   - j[I_1 S_2]_{ijk} \right] \\
& + \beta_2 (k + 1) \overline{\sigma}_2^I   \left( \Theta_{b,2}^I  [I_1S_2]_{(i-1)j(k+1)} + \overline{\Theta}_{b,2}^I  [I_1S_2]_{ij(k+1)} \right) - \beta_2 \Theta_{b,2} k [I_1S_2]_{ijk}
\end{aligned}\\
&\begin{aligned}
\frac{d}{dt}[I_1 I_2]_{ijk} = &  - \alpha_1 [I_1 I_2]_{ijk} + \beta_1 \Theta_1 (i  +  1)\sigma_1^I [S_1 I_2]_{(i+1)jk} + \beta_1(1  +   \Theta_1)   \left[ (i  +  1)[I_1I_2]_{(i+1)jk}   - i[I_1 I_2]_{ijk} \right] \\ 
& + \beta_1 (k  +  1)\sigma_1^I   \left( \Theta_{b,1}^I [S_1I_2]_{i(j-1)(k+1)} + \overline{\Theta}_{b,1}^I [S_1I_2]_{ij(k+1)} \right) \\ 
& + \beta_1 (k  +  1) \left[ \Big( \Phi_{b,1}^I   +   \Theta_{b,1}^I\Big) [I_1I_2]_{i(j-1)(k+1)} + \Big( \overline{\Phi}_{b,1}^I  +   \overline{\Theta}_{b,1}^I \Big) [I_1I_2]_{ij(k+1)} \right]  - \beta_1(1+\Theta_{b,1})k[I_1I_2]_{ijk} \\
& - \alpha_2 [I_1 I_2]_{ijk} + \beta_2 \Theta_2 (j  +  1)\sigma_2^I [I_1 S_2]_{i(j+1)k} + \beta_2(1  +   \Theta_2)   \left[ (j  +  1)[I_1I_2]_{i(j+1)k}   - j[I_1 I_2]_{ijk} \right] \\
& + \beta_2 (k  +  1)\sigma_2^I   \left( \Theta_{b,2}^I [I_1S_2]_{(i-1)j(k+1)} + \overline{\Theta}_{b,2}^I [I_1S_2]_{ij(k+1)} \right) \\ 
& + \beta_2 (k  +  1) \left[ \Big( \Phi_{b,2}^I   +   \Theta_{b,2}^I\Big) [I_1I_2]_{(i-1)j(k+1)} + \Big( \overline{\Phi}_{b,2}^I  +   \overline{\Theta}_{b,2}^I \Big) [I_1I_2]_{ij(k+1)} \right]  - \beta_2(1+\Theta_{b,2})k[I_1I_2]_{ijk} 
\end{aligned}\\
&\begin{aligned}
\frac{d}{dt}[I_1 R_2]_{ijk} = &  - \alpha_1 [I_1 R_2]_{ijk} + \beta_1 \Theta_1 (i  +  1)\sigma_1^R [S_1 R_2]_{(i+1)jk} + \beta_1(1  +   \Theta_1)   \left[ (i  +  1)[I_1R_2]_{(i+1)jk}   - i[I_1 R_2]_{ijk} \right] \\ 
& + \beta_1 (k  +  1)\sigma_1^R   \left( \Theta_{b,1}^R [S_1R_2]_{i(j-1)(k+1)} + \overline{\Theta}_{b,1}^R [S_1R_2]_{ij(k+1)} \right) \\ 
& + \beta_1 (k  +  1) \left[ \Big( \Phi_{b,1}^R   +   \Theta_{b,1}^R\Big) [I_1R_2]_{i(j-1)(k+1)} + \Big( \overline{\Phi}_{b,1}^R  +   \overline{\Theta}_{b,1}^R \Big) [I_1R_2]_{ij(k+1)} \right]  - \beta_1(1+\Theta_{b,1})k[I_1R_2]_{ijk} \\
& + \alpha_2 [I_1 I_2]_{ijk} + \beta_2 \Theta_2 \left[ (j  +  1)[I_1R_2]_{i(j+1)k}   - j[I_1 R_2]_{ijk} \right] \\
& + \beta_2 (k  +  1) \left( \Theta_{b,2}^I  [I_1R_2]_{(i-1)j(k+1)} +\overline{\Theta}_{b,2}^I   [I_1R_2]_{ij(k+1)} \right)  - \beta_2\Theta_{b,2}k[I_1R_2]_{ijk}
\end{aligned}\\
\end{align}
\newpage
\begin{align}
&\begin{aligned}
\frac{d}{dt}[R_1 S_2]_{ijk} = & \ \alpha_1 [I_1 S_2]_{ijk} + \beta_1 \Theta_1 \left[ (i  +  1)[R_1S_2]_{(i+1)jk}   - i[R_1 S_2]_{ijk} \right] \\
& + \beta_1 (k  +  1) \left( \Theta_{b,1}^S  [R_1S_2]_{i(j-1)(k+1)} +\overline{\Theta}_{b,1}^S   [R_1S_2]_{ij(k+1)} \right) 
 - \beta_1\Theta_{b,1}k[R_1S_2]_{ijk} \\
& + \beta_2 \Theta_2   \left[ (j  +  1)\overline{\sigma}_2^R   [R_1 S_2]_{i(j+1)k}   - j[R_1 S_2]_{ijk} \right] \\
& + \beta_2 (k + 1) \overline{\sigma}_2^R   \left( \Theta_{b,2}^R  [R_1S_2]_{(i-1)j(k+1)} + \overline{\Theta}_{b,2}^R  [R_1S_2]_{ij(k+1)} \right) - \beta_2 \Theta_{b,2} k [R_1S_2]_{ijk}
\end{aligned}\\
&\begin{aligned}
\frac{d}{dt}[R_1 I_2]_{ijk} = &  \ \alpha_1 [I_1 I_2]_{ijk} +  \beta_1 \Theta_1 \left[ (i  +  1)[R_1I_2]_{(i+1)jk}   - i[R_1 I_2]_{ijk} \right] \\ 
& + \beta_1 (k  +  1) \left( \Theta_{b,1}^I  [R_1I_2]_{i(j-1)(k+1)} +\overline{\Theta}_{b,1}^I   [R_1I_2]_{ij(k+1)} \right) 
 - \beta_1\Theta_{b,1}k[R_1I_2]_{ijk} \\
& - \alpha_2 [R_1 I_2]_{ijk} + \beta_2 \Theta_2 (j  +  1)\sigma_2^R [R_1 S_2]_{i(j+1)k} + \beta_2(1  +   \Theta_2)   \left[ (j  +  1)[R_1I_2]_{i(j+1)k}   - j[R_1 I_2]_{ijk} \right] \\
& + \beta_2 (k  +  1)\sigma_2^R   \left( \Theta_{b,2}^R [R_1S_2]_{(i-1)j(k+1)} + \overline{\Theta}_{b,2}^R [R_1S_2]_{ij(k+1)} \right) \\ 
& + \beta_2 (k  +  1) \left[ \Big( \Phi_{b,2}^R   +   \Theta_{b,2}^R\Big) [R_1I_2]_{(i-1)j(k+1)} + \Big( \overline{\Phi}_{b,2}^R  +   \overline{\Theta}_{b,2}^R \Big) [R_1I_2]_{ij(k+1)} \right]  - \beta_2(1+\Theta_{b,2})k[R_1I_2]_{ijk} 
\end{aligned}\\
&\begin{aligned}
\frac{d}{dt}[R_1 R_2]_{ijk} = & \ \alpha_1 [I_1 R_2]_{ijk} + \beta_1 \Theta_1 \left[ (i  +  1)[R_1R_2]_{(i+1)jk}   - i[R_1 R_2]_{ijk} \right] \\ 
& + \beta_1 (k  +  1) \left( \Theta_{b,1}^R  [R_1R_2]_{i(j-1)(k+1)} +\overline{\Theta}_{b,1}^R   [R_1R_2]_{ij(k+1)} \right) 
 - \beta_1\Theta_{b,1}k[R_1R_2]_{ijk} \\
& + \alpha_2 [R_1 I_2]_{ijk} + \beta_2 \Theta_2 \left[ (j  +  1)[R_1R_2]_{i(j+1)k}   - j[R_1 R_2]_{ijk} \right] \\
& + \beta_2 (k  +  1) \left( \Theta_{b,2}^R  [R_1R_2]_{(i-1)j(k+1)} +\overline{\Theta}_{b,2}^R   [R_1R_2]_{ij(k+1)} \right)  - \beta_2\Theta_{b,2}k[R_1R_2]_{ijk} \ .
\end{aligned}
\end{align}
Once again, it is possible to verify that this system of ODEs satisfy the conservation of nodes,  
\begin{align}
\sum_{X, Y} \sum_{i,j,k} \frac{d}{dt} [X_1 Y_2]_{ijk} = 0 \ .
\end{align}
The dynamics of agent 1 is initialized at time $t=0$ with the following initial conditions:
\begin{equation}
\begin{array}{@{\extracolsep{40pt}} l l l }
\ [S_1 S_2]_{ijk}(0) = (1-\epsilon_1) \rho(i,j,k)  & [I_1 S_2]_{ijk}(0) = \epsilon_1 \rho(i,j,k)  & [R_1 S_2]_{ijk}(0) = 0 \\
\ [S_1 I_2]_{ijk}(0) = 0  & [I_1 I_2]_{ijk}(0) = 0  & [R_1 I_2]_{ijk}(0) = 0 \\
\ [S_1 R_2]_{ijk}(0) = 0  & [I_1 R_2]_{ijk}(0) = 0  & [R_1 R_2]_{ijk}(0) = 0 \ .
\end{array}
\end{equation}
The dynamics of agent 2 is initialized at time $t=\tau \geq 0$ by making the following substitutions:
\begin{equation}
\begin{array}{@{\extracolsep{15pt}} l l l }
\ [S_1 S_2]_{ijk}(\tau) \to (1-\epsilon_2) [S_1 S_2]_{ijk}(\tau)  & [I_1 S_2]_{ijk}(\tau) \to (1-\epsilon_2) [I_1 S_2]_{ijk}(\tau)  & [R_1 S_2]_{ijk}(\tau) \to (1-\epsilon_2) [R_1 S_2]_{ijk}(\tau) \\
\ [S_1 I_2]_{ijk}(\tau) \to \epsilon_2 [S_1 S_2]_{ijk}(\tau)   & [I_1 I_2]_{ijk}(\tau) \to \epsilon_2 [I_1 S_2]_{ijk}(\tau)   & [R_1 I_2]_{ijk}(\tau) \to \epsilon_2 [R_1 S_2]_{ijk}(\tau)  \ .
\end{array}
\end{equation}

\end{document}